\documentclass{article}
\usepackage{jheppub}
\usepackage{xcolor}
\usepackage{mathtools}
\DeclarePairedDelimiter\abs{\lvert}{\rvert}%

\title{Uncool soft-wall transitions and gravitational waves}
\author[a,b,c]{Ameen Ismail,}
\author[a,b,c,d,e]{Lian-Tao Wang}
\affiliation[a]{Department of Physics, University of Chicago, Chicago, IL 60637, USA}
\affiliation[b]{Enrico Fermi Institute, University of Chicago, Chicago, IL 60637, USA}
\affiliation[c]{Leinweber Institute for Theoretical Physics, University of Chicago, Chicago, IL 60637, USA}
\affiliation[d]{Kavli Institute for Cosmological Physics, University of Chicago, Chicago, IL 60637, USA}
\affiliation[e]{HEP Division, Argonne National Laboratory, 9700 Cass Ave., Argonne, IL 60439, USA}
\emailAdd{ameenismail@uchicago.edu}
\emailAdd{liantaow@uchicago.edu}

\abstract{ 
Theories with warped extra dimensions, like the Randall--Sundrum (RS) model, exhibit a holographic phase transition from a hot, deconfined black brane phase to a cool, confined phase.
The standard picture of a first-order, strongly supercooled phase transition is expected to change in variations where the extra dimension is smoothly cut off by a soft-wall curvature singularity, as opposed to a hard brane.
To understand this situation, we consider a simple ansatz for the warped geometry which allows us to obtain analytical results while maintaining the essential behavior of a soft wall.
Unlike RS with the usual Goldberger--Wise stabilization, the hot, black brane phase only exists above a minimum temperature, which is not much smaller than the critical temperature.
We explore the dynamics of the phase transition across the range of possibilities for the asymptotic geometry of a soft wall.
This involves calculating an effective 4D action for the location of the black brane horizon.
Using the effective action, we show that the phase transition completes rapidly ($\beta/H$ of $10^{3\text{--}4}$ is typical) and with only slight supercooling. We compute the resulting gravitational wave signal for a TeV-scale transition, finding that it is accessible to future space-based interferometers. 
}
\begin{document}

\maketitle
\flushbottom

\section{Introduction}\label{sec:intro}

Models of warped extra dimensions are widespread in particle physics beyond the Standard Model. Among other applications, they provide a mechanism to stabilize a large hierarchy between the electroweak scale and the Planck scale, and thus have the potential to address the Higgs naturalness problem. The basic concept as proposed in the Randall--Sundrum (RS) model features a slice of 5D anti-de Sitter (AdS) space capped by UV and IR branes~\cite{Randall:1999ee}. RS and related models can also be analyzed in the AdS/CFT correspondence~\cite{Maldacena:1997re,Witten:1998qj,Gubser:1998bc}. They are dual to strongly-coupled, near-conformal sectors, with the RS solution to the hierarchy problem dual to Higgs compositeness~\cite{Arkani-Hamed:2000ijo,Rattazzi:2000hs}. At a scale corresponding to the location of the IR brane the symmetry is spontaneously broken and the CFT confines. Fields localized toward the IR and the UV are respectively dual to composites of the CFT and elementary fields.

At high temperatures, one expects that the conformal symmetry is restored. In the 5D picture this corresponds to the IR brane being hidden behind a black brane horizon. As the temperature is lowered a phase transition (PT) to the confined phase takes place via the nucleation of bubbles of IR brane within the black brane background. The conformal PT in warped models has been the subject of intense study~\cite{Creminelli:2001th,Randall:2006py,Konstandin:2010cd,Konstandin:2011dr,vonHarling:2017yew,Bruggisser:2018mus,Bruggisser:2018mrt,Baratella:2018pxi,Agashe:2019lhy,Agashe:2020lfz,Agrawal:2021alq,Baldes:2021aph,Bruggisser:2022rdm,Csaki:2023pwy,Eroncel:2023uqf,Megias:2023kiy,Ferrante:2023bcz,Mishra:2023kiu,Luo:2025alo,Agrawal:2025wvf,Mishra:2026lvq}. Provided the IR brane lies in a region where the geometry is approximately AdS, one can analyze the PT within the dilaton effective field theory.
Importantly, the PT may be first-order and strongly supercooled, particularly in the case when the size of the extra dimension is stabilized through the Goldberger--Wise mechanism~\cite{Goldberger:1999uk}. This is interesting from an experimental perspective, as it leads a large stochastic gravitational wave (GW) background observable at next-generation detectors.

The truncation of the extra dimension by an IR brane corresponds to a limit in which the conformal symmetry is broken by an operator with infinite scaling dimension~\cite{Rattazzi:2000hs}. This is not the only mechanism by which confinement can take place. The other possibility is that the extra dimension is smoothly cut off by the appearance of a curvature singularity in the bulk (a ``soft wall'', as opposed to the ``hard wall'' IR brane).
The PT is harder to study in this scenario, as the geometry near the singularity is strongly deformed away from AdS by the backreaction to a bulk field. However, it is necessary to analyze the conformal PT at strong backreaction to have a comprehensive understanding of it. The authors of~\cite{Mishra:2024ehr}, focusing on a geometry inspired by constructions in string theory, observed major differences with the standard RS PT lore. They find two branches of black branes: at a given temperature, there are two black brane solutions, one of which is stable and the other of which is unstable. Moreover, the black holes only exist above a minimum temperature $T_{\rm min}$ which is not much smaller than the critical temperature $T_c$. This suggests that the PT cannot be strongly supercooled, which would certainly affect the expected GW signal. It was argued in~\cite{Gursoy:2008za} that these are general features of soft walls; they are also exhibited by some examples from string theory, including Klebanov--Strassler and Klebanov--Witten black holes~\cite{Buchel:2021yay,Buchel:2018bzp,Buchel:2024phy}. 

The study of the PT dynamics and associated GW signals is lacking in soft wall models\footnote{See also~\cite{Bea:2021zol} for a novel production mechanism for GWs involving the unstable black brane branch.}.
The bounce action is dominated by the hot phase, in which the dilaton effective theory is not applicable.
There are a couple of works that focus on specific holographic QCD models~\cite{Bigazzi:2020avc,Morgante:2022zvc}. Some authors have studied the PT with particular polynomial potentials for the bulk stabilizing field, also finding a minimum black brane temperature~\cite{Ares:2020lbt,Ares:2021nap,Ares:2021ntv}.
However, these geometries do not describe a theory that confines, but rather exhibit a PT between two different deconfined CFTs.
Here we are interested specifically in soft walls that trigger confinement. This generally requires a potential that grows exponentially in the bulk field~\cite{Gursoy:2007cb,Gursoy:2007er}, rather than the polynomial behavior considered in~\cite{Ares:2020lbt,Ares:2021nap,Ares:2021ntv}.

The goal of this work is to better understand the confinement transition in such warped geometries with strong backreaction. We review these solutions to the 5D Einstein equations in Section~\ref{sec:warped}. We emphasize that the asymptotic behavior near the curvature singularity is tied to confinement. Confinement requires the potential\footnote{Actually the superpotential $W$ defined in Section~\ref{sec:warped}, which is related to the potential.} for a bulk scalar field to grow exponentially, and the near-singularity geometry is determined by the exponential growth rate. This can be parametrized by a dimensionless quantity $\nu$ such that confinement occurs for $1 \leq \nu < 2$. As $\nu$ increases the wall becomes ``harder'', with the solution being pathological for $\nu > 2$. We then study the system at finite temperature and discuss some simple cases (a constant and an exponential potential) in which one can obtain analytical solutions. We also review the equilibrium thermodynamics of the black brane solutions.

In Section~\ref{sec:pt_setup} we model the PT as a single-field bounce, where the horizon location is a function of the radial coordinate of a bubble. The bubbles interpolate between the confined solution, with the horizon at the singularity, and the deconfined solution, with the horizon at a location corresponding to the temperature of the thermal bath. This approximation reduces the problem to studying the bounce configuration of a single field (the horizon location); we show how to compute the effective action for the horizon.

Our main results are presented in Section~\ref{sec:main_result}.
We consider a simple ansatz for the metric which stitches together the solutions for a constant potential and an exponential potential. This is the simplest possibility that has the correct limiting behavior in the UV and IR to capture the salient features of a soft-wall confinement PT. In terms of the exponential growth rate $\nu$, this ansatz describes all confining soft walls except the edge case $\nu = 1$.
Using this metric, we are able to obtain analytical results for thermodynamic quantities and the effective action. As expected, we find two black hole branches above a minimum temperature, one of which is stable. We compute the bounce action and the nucleation temperature for the PT, finding that the PT completes without much supercooling. We then study the GW signals produced in the PT, focusing on a few benchmark points. Despite the weaker PT, we find that a TeV-scale transition is observable at future space-based gravitational detectors like AEDGE~\cite{AEDGE:2019nxb,Badurina:2019hst}, BBO~\cite{Crowder:2005nr,Corbin:2005ny,Harry:2006fi}, and DECIGO~\cite{Seto:2001qf,Kawamura:2011zz,Yagi:2011wg,Isoyama:2018rjb}. In optimistic scenarios, a $\sim 100$~GeV PT could be seen at LISA~\cite{LISA:2017pwj,Baker:2019nia}, while a $\sim 100$~TeV PT could be probed by terrestrial interferometers like the Cosmic Explorer~\cite{LIGOScientific:2016wof,Reitze:2019iox} and Einstein Telescope~\cite{Punturo:2010zz,Hild:2010id,Sathyaprakash:2012jk,ET:2019dnz}.

We also study the edge case, corresponding to $\nu = 1$, in Section~\ref{sec:linear_dilaton}. We must include a subleading term in the near-singularity geometry to understand the PT dynamics. This also makes it infeasible to find analytical expressions for thermodynamic quantities, except for a special case corresponding to a linear dilaton geometry. We therefore compute the phase diagram and GW signals numerically. We find the same qualitative behavior as the soft walls treated in Section~\ref{sec:main_result}, again with the exception of a linear dilaton geometry. We study the latter analytically to show that the PT is second order and that there is only one black brane branch. This is relevant to works that have employed such geometries to study models with a gapped continuum spectrum
~\cite{Falkowski:2008fz,Falkowski:2008yr,Cabrer:2009we,Bellazzini:2015cgj,Csaki:2018kxb,Megias:2019vdb,Megias:2021mgj,Csaki:2021gfm,Csaki:2021xpy,Csaki:2022lnq,Fichet:2022xol,Fichet:2023dju,Fichet:2023xbu,Ferrante:2023fpx,Ferrante:2025ofe}.
We summarize our findings in Section~\ref{sec:conclusions}.

\section{Warped constructions and soft walls}\label{sec:warped}

\subsection{Zero-temperature phase}
We first discuss some preliminaries regarding warped geometries, beginning with the the cold, zero-temperature phase.

We consider the following 5D action involving a stabilizing scalar field $\phi$:
\begin{equation}
    S = \int d^5 x \sqrt{g} \left[ -\frac{1}{2\kappa^2} R + \frac{1}{2} g^{AB} \partial_A \phi \partial_B \phi - V(\phi) \right] .
\end{equation}
Here $\kappa$ is related to the 5D Planck scale $M_5$ as $1/\kappa^2 = 4 M_5^3$.

For the 5D metric, we take the ansatz
\begin{equation}\label{eq:metric_ansatz_cold}
    ds^2 = e^{-2A(y)} \left[ dt^2 - dx^i dx_i \right] - dy^2 ,
\end{equation}
where $A(y)$ is the warp factor.

One can put the Einstein equations in the following form, where the primes denote $y$-derivatives:
\begin{align}
    \kappa^2 \phi'^2 &= 3 A'' \\
    \kappa^2 V(\phi(y)) &= -6 A'^2 + \frac{3}{2} A'' .
\end{align}
Note that the scalar equation of motion is implied by the Einstein equations.
For a constant potential $V(\phi) = -6k^2/\kappa^2$ , one obtains the AdS solution $A = k y$. 

One can generate solutions to these equations with the superpotential method~\cite{DeWolfe:1999cp,Csaki:2000zn}. Given a superpotential function $W[\phi]$, a solution is given by
\begin{align}
    A' &= \frac{\kappa^2}{6} W[\phi] \\
    \phi' &= \frac{1}{2} \frac{d W[\phi]}{d\phi} \\
    V(\phi) &= \frac{1}{8} \left( \frac{d W[\phi]}{d\phi} \right)^2 - \frac{\kappa^2}{6} W[\phi]^2 .
\end{align}
When one has UV/IR branes, there are also boundary terms in these equations. In this work, though, we will not have an IR brane, and we ignore the effect of the UV brane.
For a constant superpotential $W = 6k/\kappa^2$, one obtains AdS. This method is convenient because it provides an exact solution to the 5D Einstein equations, fully  incorporating the backreaction of the scalar field on the metric. In particular, we are interested in cases where the asymptotic behavior of the superpotential is exponentially growing in $\phi$; this can lead to soft-wall confinement in the 5D geometry.

As a concrete illustration of these ideas, we consider the superpotential introduced in~\cite{Cabrer:2009we},
\begin{equation}\label{eq:quiros_w}
    W = \frac{6k}{\kappa^2}(1 + \exp \nu \kappa \phi / \sqrt{3} ) ,
\end{equation}
where $0 < \nu < 2$ is a free parameter\footnote{The requirement $\nu < 2$ is imposed to that the singularity that develops in the extra dimension is a ``good'' one in the sense of Gubser's criterion~\cite{Gubser:2000nd}.}. Using the superpotential equations, we find that the warp factor and scalar profile are given by
\begin{align}\label{eq:quiros_metric}
    \frac{\kappa}{\sqrt{3}} \phi &= - \frac{1}{\nu} \log \left[ \nu^2 k(y_s - y) \right] \\
    A(y) &= k y - \frac{1}{\nu^2} \log \left( 1 - \frac{y}{y_s} \right) .
\end{align}
This solution is asymptotically AdS in the deep UV, $y \rightarrow - \infty$.
As $y \rightarrow y_s$ the warp factor diverges and the geometry is cut off by a singularity. Note that in the near-singularity region the scalar profile backreacts strongly on the metric, deforming it away from AdS. By studying the spectrum, one can show that this geometry leads to confinement for $\nu \geq 1$.

The underlying reason that the spectrum changes at $\nu = 1$ is the behavior of $dA/dz$ near the singularity, where $z$ is the conformal coordinate defined by $dz/dy = e^A$. Near the singularity $dA/dz = e^{-A} A' \sim  e^{-k y} (y_s - y)^{1/\nu^2 - 1}$. Hence $dA/dz$ goes to $0$ or $\infty$ at the singularity, depending on whether $\nu$ is larger or smaller than $1$. Larger values of $\nu$ correspond to ``harder'' walls, in the sense that $dA/dz$ diverges more quickly near the singularity.

More generally, we can classify confining backgrounds by the asymptotic behavior of the superpotential at large $\phi$. A superpotential that grows as
\begin{equation}
    W \sim \phi^n \exp \kappa \nu \phi / \sqrt{3}
\end{equation}
confines for $n \geq 0$ and $1 \leq \nu < 2$~\cite{Gursoy:2007cb,Gursoy:2007er}. The warp factor (and the scalar profile) near the singularity are also determined by the asymptotics of the superpotential. As $y \rightarrow y_s$ the leading behavior is
\begin{equation}\label{eq:IR_asymptotics}
    A \sim -\frac{1}{\nu^2} \log(1 - y/y_s) ,
\end{equation}
for all $n, \nu$. In the special case $\nu = 1$, the subleading behavior matters as well. We will briefly discuss this special case in Section~\ref{sec:linear_dilaton}. For most of this paper we will be interested in geometries with the asymptotic behavior in Eq.~\eqref{eq:IR_asymptotics}.

\subsection{Hot phase}
Let us now consider the hot phase of the theory. At finite temperature we expect a black brane horizon to appear at $y = y_h$. We modify our metric ansatz to include a blackening factor $b(y)$,
\begin{equation}\label{eq:metric_ansatz_hot}
    ds^2 = e^{-2A(y)} \left[ b(y) dt^2 - dx^i dx_i \right] - \frac{dy^2}{b(y)} .
\end{equation}
The blackening factor vanishes at the horizon (and it approaches $1$ far away from the horizon). The Einstein equations may be written as follows:
\begin{align}\label{eq:einstein_equations_hot}
    b'' &= 4 A' b' \\
    \kappa^2 \phi'^2 &= 3 A'' \\
    2 \kappa^2 \frac{V(\phi)}{b} &= -12 A'^2 + 3 A' \frac{b'}{b} + \kappa^2 \phi'^2 .
\end{align}

Finding analytical solutions in the hot phase is much harder than at zero temperature. However, in the special cases of a constant potential and an exponential potential, we can find a solution. For an asymptotically AdS extra dimension with a confining soft wall (like Eq.~\eqref{eq:quiros_metric}), we expect that these special cases describe the UV and IR limits of the geometry at finite temperature.

For a constant potential $V = -6k^2/\kappa^2$, the solution is AdS-Schwarzschild: $A = k y$, $\phi = {\rm constant}$, $b = 1 - \exp 4 k (y - y_h)$. Note that in this case $A$ and $\phi$ are the same at finite temperature and at zero temperature---only the blackening factor changes. This is not a generic feature and it is related to why we can obtain a simple analytical solution.

The same thing happens for an exponential potential. For concreteness, let us consider the large-$\phi$ limit of the potential resulting from Eq.~\eqref{eq:quiros_w},
\begin{equation}
    V(\phi) = -\frac{6k^2}{\kappa^2} \left( 1 - \frac{\nu^2}{4} \right) e^{2 \nu \kappa \phi / \sqrt{3}} .
\end{equation}
A solution is given by 
\begin{align}
    \frac{\kappa}{\sqrt{3}} \phi(y) &= - \frac{1}{\nu} \log \left[ \nu^2 k(y_s - y) \right] \\
    A(y) &= - \frac{1}{\nu^2} \log \left( 1 - \frac{y}{y_s} \right) \\
    b(y) &= 1 - \left( \frac{y_s - y}{y_s - y_h} \right)^{1 - 4/\nu^2} .
\end{align}

\subsection{Thermodynamic quantities}

Given a warp factor and blackening factor, one can calculate various thermodynamic quantities associated with the black brane. We will review the computation of the temperature, the entropy, and the free energy, mainly following~\cite{Mishra:2024ehr}.

To compute the temperature, we study the metric, Eq.~\eqref{eq:metric_ansatz_hot}, near the black brane horizon. Expanding at leading order in $\epsilon = y_h - y$, we have
\begin{equation}
    -ds^2 \supset e^{-2A(y_h)} \epsilon \abs{b'(y_h)} dt_E^2 + \frac{d\epsilon^2}{\abs{b'(y_h)} \epsilon} ,
\end{equation}
working in Euclidean time $t_E = i t$. Changing variables to $r = 2 \sqrt{\epsilon / \abs{b'(y_h)}}$ and $\theta = e^{-A(y_h)} \abs{b'(y_h)} t_E / 2$, the above equation becomes the usual polar coordinate metric $dr^2 + r^2 d\theta^2$. To avoid a conical defect we require $\theta \sim \theta + 2\pi$; thus $t_E$ is periodic with period $\beta = 4 \pi e^{A(y_h)} / \abs{b'(y_h)}$. We conclude the temperature is
\begin{equation}\label{eq:temperature}
    T_h = \beta^{-1} = \frac{1}{4\pi} e^{-A(y_h)} \abs{b'(y_h)} .
\end{equation}
Again, the temperature may attain a minimum value in situations with strong backreaction, which has important ramifications for the phase transition.

We can use the Bekenstein--Hawking formula to calculate the entropy in the hot phase. The entropy density is
\begin{equation}\label{eq:entropy}
    s = \frac{A_{\rm horizon}}{4 G_N V} = 2\pi\frac{A_{\rm horizon}}{\kappa^2 V} ,
\end{equation}
where $A_{\rm horizon}$ is the area of the black brane horizon and $V$ is the 3-volume.
The metric on the horizon at $y = y_h$ is simply
\begin{equation}
    ds_{\rm horizon}^2 = e^{-2A(y_h)} dx^i dx_i ,
\end{equation}
so the area of the horizon is $A_{\rm horizon} = e^{-3A(y_h)} V$. The entropy density is thus
\begin{equation}
    s = \frac{2\pi}{\kappa^2} e^{-3A(y_h)} .
\end{equation}

Lastly, we identify the free energy density using $s = - \partial f / \partial T_h$. One has to be careful in integrating this relation to obtain $f$ because $s(T_h)$ is multi-valued when the black hole has a minimum temperature. To circumvent this issue we can work with $s(y_h)$ and $T_h(y_h)$. Then we have
\begin{equation}
    f(y_h) = f_0 - \int^{y_h} d\widetilde{y}_h s(\widetilde{y}_h) T_h'(\widetilde{y}_h) .
\end{equation}
Suppose we have a case with strong backreaction where the metric has a singularity at $y = y_s$. The cold phase corresponds to the limit in which the horizon is pushed all the way to the singularity, $y_h \rightarrow y_s$. In studying the phase transition, it is useful to choose the constant term $f_0$ such that $f$ vanishes in the cold phase. Then we have
\begin{equation}\label{eq:free_energy}
    f(y_h) = \int^{y_s}_{y_h} d\widetilde{y}_h s(\widetilde{y}_h) T_h'(\widetilde{y}_h) .
\end{equation}

\section{Effective action}\label{sec:pt_setup}

The phase transition involves the nucleation of bubbles which interpolate between the cold and hot phases. An $O(3)$-symmetric bubble should be a function of the radial coordinate $\rho = \sqrt{x^i x_i}$ and $y$. Computing the bubble requires solving Euclidean-time 5D Einstein equations, which would be difficult even numerically.

\begin{figure}
    \centering
    \includegraphics[width=\textwidth]{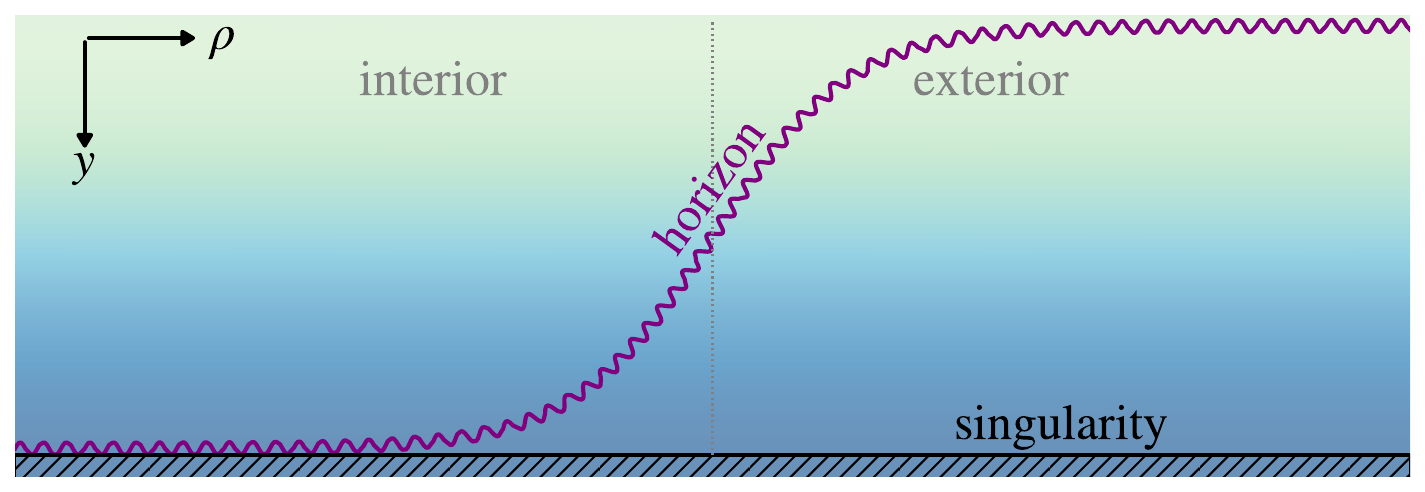}
    \caption{A schematic depiction of our parametrization of the bounce. In the interior of the bubble, the horizon is pushed all the way to the singularity, and the theory is in the cold, confined phase. In the exterior region the horizon lies at a position inside the bulk.}
    \label{fig:schematic}
\end{figure}

Instead we will approximate the bubble by assuming that the bubble at a given $\rho$ is described by an equilibrium solution with the horizon at some position $y_h(\rho)$. That is, we assume the bubble is the equilibrium solution, but with the horizon being a function of $\rho$. The same approach was taken in~\cite{Bigazzi:2020phm,Morgante:2022zvc}. We illustrate this parametrization in Fig.~\ref{fig:schematic}: the bubble interpolates between the confined phase in the interior ($\rho = 0$) and the deconfined phase in the exterior ($\rho \rightarrow \infty$).

This ansatz does not solve the Einstein equations, but it is a good approximation so long as the the bubble wall thickness $d$ is larger than the KK scale, $d M_{\rm KK} \gg 1$\footnote{The typical size of a term in the Einstein equations is $R_{\mu\nu} \sim \mathcal{O}(e^{-2ky} k^2)$, where $k$ is the inverse AdS curvature. Our bubble ansatz introduces an error of order $d^{-2}$. This is small if $d > k^{-1} e^{ky_s} \sim M_{\rm KK}^{-1}$. Since $d$ is determined by the second derivative of the effective potential, this condition is related to the usual requirement that the dilaton mass should be small compared to the KK scale to use the dilaton EFT.} . In this regime we can neglect the backreaction of the bubble on the metric. Furthermore, we expect this approximation works best at lower temperatures (further from the critical point), where the wall is thicker. This is fortunate: it means that the largest GW signals occur in the regime where our approximation is reliable.

To study the dynamics of the PT one needs to compute the effective action for the bubble. The remainder of this section is devoted to deriving the effective potential and the kinetic term. We will use these results in Section~\ref{sec:main_result} to study the PT in a simple model of soft-wall confinement.

For the potential, we consider the metric Eq.~\eqref{eq:metric_ansatz_hot} at a temperature $T \neq T_h$. Then there is a conical singularity at the horizon. Following~\cite{Creminelli:2001th}, one should regularize this with a spherical cap to compute the contribution of the conical defect to the free energy. The result is that the effective potential is given by
\begin{equation}\label{eq:eff_potential_general}
    V(y_h) = f(y_h) - s(y_h) \left[ T - T_h(y_h) \right] .
\end{equation}

We can derive the kinetic term by considering a perturbation of the horizon location. This was done for AdS-Schwarzschild in~\cite{Bigazzi:2020phm}. We parametrize the perturbation by appropriately modifying the metric ansatz Eq.~\eqref{eq:metric_ansatz_hot}:
\begin{equation}
    ds^2 = e^{-2A(y + r(\vec{x}))} \left( b(y) dt^2 - dx^i dx_i \right) - \frac{1}{b(y)} dy^2 .
\end{equation}
The reason we only need to modify $A(y)$ and not $b(y)$ is that taking $A(y) \rightarrow A(y+\Delta y)$, $b(y) \rightarrow b(y + \Delta y)$ corresponds to an unphysical shift in the definition of $y$.
Intuitively, the physical quantity is the warp factor evaluated at the horizon.

We now expand the action to two derivatives and extract the pieces which are proportional to $(\nabla r)^2$. These terms arise from the bulk Ricci scalar and the kinetic term for the bulk scalar $\phi$.
From the Ricci scalar we find a contribution to the kinetic term
\begin{equation}
    -\frac{1}{2\kappa^2} \int d^5 x \sqrt{g} R \supset \frac{3}{\kappa^2} \int d^5 x e^{-2A(y+r)}\left(-A'(y + r)^2 (\nabla r)^2 + A''(y + r) (\nabla r)^2 + A'(y+r) \nabla^2 r \right)
\end{equation}
where we retained only those terms with two derivatives of $r$. This can be simplified by adding a total derivative $\nabla ( e^{-2A} A' \nabla r )$, leading to a contribution
\begin{equation}\label{eq:kinetic_ricci}
    \frac{3}{\kappa^2} \int d^5 x e^{-2A(y+r)}A'(y + r)^2 (\nabla r)^2 .
\end{equation}

From the kinetic term for the bulk scalar we get a contribution
\begin{equation}
    \int d^5 x \frac{1}{2} \sqrt{g} g^{AB} \partial_A \phi(y + r(\vec{x})) \partial_B \phi(y + r(\vec{x})) \supset - \int d^5 x \frac{1}{2} e^{-2A(y + r)} \phi'(y + r)^2 (\nabla r)^2 ,
\end{equation}
again retaining only those terms with two derivatives of $r(\vec{x})$. Using the equation of motion $\kappa^2 \phi'^2 = 3A''$, we can rewrite this as
\begin{equation}\label{eq:kinetic_scalar}
    -\frac{3}{2\kappa^2} \int d^5 x e^{-2A(y+r)} A''(y+r) (\nabla r)^2 .
\end{equation}
Adding Eqs.~\eqref{eq:kinetic_ricci} and~\eqref{eq:kinetic_scalar} we obtain the kinetic term,
\begin{equation}\label{eq:kinetic_combined}
\begin{split}
    &\frac{3}{\kappa^2} \int d^5 x e^{-2A(y + r)} \left[ A'(y+r)^2 - \frac{1}{2} A''(y+r) \right] (\nabla r)^2  \\
    &= -\frac{3}{2\kappa^2} \int d^5 x \partial_y \left[ e^{-2A(y+r)} A'(y+r) (\nabla r)^2 \right]
\end{split}    
\end{equation}
The overall negative sign is because we are working in Lorentzian signature.

The lower limit in the $y$-integral in Eq.~\eqref{eq:kinetic_combined} diverges because we have sent the UV brane to the AdS boundary. By restoring the UV brane and introducing a brane-localized counterterm we can cancel the divergence~\cite{Skenderis:2002wp,Bigazzi:2020phm}. This leads to our final result for the kinetic term,
\begin{equation}\label{eq:kinetic_final}
    \frac{3}{2\kappa^2} \int d^4 x e^{-2A(y)} A'(y) (\nabla r)^2 \Big |_{y = y_h}.
\end{equation}
The procedure we followed here is similar to how one derives the dilaton kinetic term in RS by treating the IR brane location as $x^\mu$-dependent~\cite{Csaki:1999mp,Goldberger:1999un}. For an AdS-Schwarzschild geometry we reproduce the kinetic term derived in~\cite{Bigazzi:2020phm}.

\section{A soft-wall case study}\label{sec:main_result}

\subsection{Setup}

We would like to study the conformal PT in soft-wall geometries like those described by Eq.~\eqref{eq:IR_asymptotics}, corresponding to a superpotential that grows as $W \sim \exp \nu \kappa \phi / \sqrt{3}$. Recall that these give rise to confinement and a gapped spectrum from $1 \leq \nu < 2$, with the value of $\nu$ determining the near-singularity geometry (i.e. the ``hardness'' of the wall). Here we are specifically interested in $1 < \nu < 2$; we will discuss the edge case $\nu = 1$ in Section~\ref{sec:linear_dilaton}. This encompasses many interesting cases, including the string-inspired geometry studied in~\cite{Mishra:2024ehr}.

This endeavor is complicated by the fact that we do not have many analytical solutions in the black brane phase.
However, we know that the UV and IR limits are described by the solutions for a constant and exponential potential, respectively.
This suggests we make the piecewise approximation
\begin{equation}\label{eq:warp_factor_deriv}
    A' = \begin{cases}
        k & y < y_i \equiv y_s - 1/k\nu^2 \\
        \frac{1}{\nu^2(y_s - y)} & y > y_i .
    \end{cases}
\end{equation}
Essentially, we take the warp factors for a constant and exponential potential and stitch them together at $y = y_i$, such that $A'$ is continuous. This parametrizes a family of soft walls that become ``harder'' as $\nu$ increases.

The merit of this ansatz is that it has the correct behavior in the UV and IR to capture the dynamics of the PT in soft-wall confinement, while still being simple enough to study analytically.
We are able to obtain analytical expressions for the blackening factor, thermodynamic quantities, and the effective action for the black brane horizon.
In what follows we present results for $\nu = \sqrt{2}$, which simplifies the computations. This is sufficient to demonstrate the essence of the PT dynamics. When we study the GW signals associated with the PT in section~\ref{sec:GW}, we will consider benchmarks at a variety of $\nu$.

We find the warp factor by integrating Eq.~\eqref{eq:warp_factor_deriv}:
\begin{equation}
    A(y) = \begin{cases}
        k y & y < y_i \\
        k y_s - \frac{1}{2} - \frac{1}{2} \log 2 k (y_s - y) & y > y_i .
    \end{cases}
\end{equation}
The blackening factor which solves the Einstein equations, Eq.~\eqref{eq:einstein_equations_hot} is given by
\begin{equation}
    1 - b(y) = \begin{cases}
        e^{4k(y - y_h)} & y, y_h < y_i \\
        \frac{y_s - y_h}{y_s - y} \frac{k(y_s - y) - 1}{k(y_s - y_h) - 1} & y, y_h > y_i .
    \end{cases}
\end{equation}
When $y_h > y_i$, the blackening factor in the AdS region $y < y_i$ is just given by the AdS-Schwarzschild result up to a constant, $1 - b(y) = c e^{4 k y}$. The constant is determined by matching at $y_i$. The precise form is not important for deriving thermodynamic quantities.

Using Eq.~\eqref{eq:temperature} it is easy to find the temperature:
\begin{equation}\label{eq:temperature_soft_wall}
    \pi \frac{T_h}{k} = \begin{cases}
        e^{-k y_h} & y_h < y_i \\
        e^{-k y_i} \frac{\sqrt{2\psi}}{4 \psi - 4 \psi^2} & y_h > y_i
    \end{cases} ,
\end{equation}
where $\psi = k(y_s - y_h)$. This shows that the temperature attains a minimum at $y_h = y_s - 1/3k$ of
\begin{equation}
    T_{\rm min} = \left( \frac{3}{2} \right)^{3/2} \frac{ k e^{-k y_s + 1/2}}{2\pi}
\end{equation}
and goes to infinity as $y_h \rightarrow y_s$, which is the behavior we expected. The entropy and free energy also follow from Eqs.~\eqref{eq:entropy} and~\eqref{eq:free_energy}; the latter is given by
\begin{equation}\label{eq:free_energy_soft_wall}
    2\frac{\kappa^2}{k} f(y_h) = \begin{cases}
        -e^{-4 k y_h} + 3 (2 \log 2 - 1) e^{-4 y_i} & y < y_i \\
        2 e^{-4 y_i} \frac{(2\psi +3 (1 - \psi) \log (1 - \psi))}{\psi-1} & y > y_i .
    \end{cases}
\end{equation}
We choose the overall constant such that the free energy vanishes in the confined phase, $f(y_s) = 0$. The free energy also vanishes at the critical temperature $T_c$, given by
\begin{equation}
    T_c = \left[3 (2 \log 2 - 1) \right]^{1/4} \frac{k e^{-k y_s + 1/2}}{\pi} .
\end{equation}
Note that $T_{\rm min} / T_c \approx 0.9$, suggesting that not much supercooling is possible. We remark that $T_{\rm min} / T_c$ is a function of $\nu$ (which we have been fixing to be $\sqrt{2}$). As $\nu \rightarrow 1$, $T_{\rm min}/ T_c \rightarrow 1$, and as $\nu \rightarrow 2$, $T_{\rm min}/T_c \rightarrow 0$. To achieve a strongly supercooled PT requires tuning $\nu$ close to 2; for instance, to obtain $T_{\rm min} / T_c = 0.5$ requires $\nu \approx 1.99$.

We show the phase diagram in the left panel of Fig.~\ref{fig:free_energy_potential}. It exhibits the characteristic ``shark-fin'' structure also seen in~\cite{Gursoy:2008za,Buchel:2021yay,Mishra:2024ehr}. Above the minimum temperature $T_{\rm min}$, there exist two black brane solutions distinguished by the horizon location $y_h$. The solution with smaller $y_h$ is thermodynamically favored over the one with larger $y_h$. For $T_{\rm min} < T < T_c$, the free energy of the black brane is positive, indicating it is metastable. The two branches of black branes coincide at $T_{\rm min}$; below $T_{\rm min}$ only the confined phase exists.

\subsection{Bounce action}

\begin{figure}
    \centering
    \includegraphics[width=\textwidth]{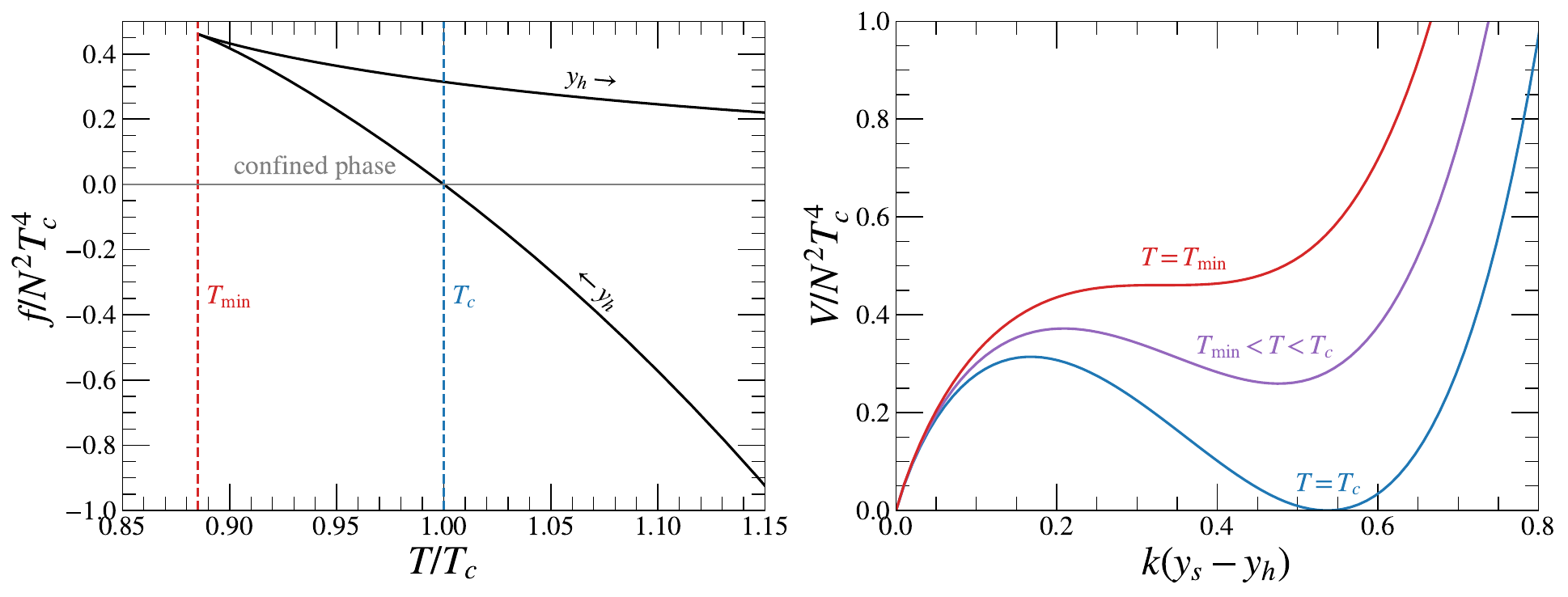}
    \caption{\textbf{Left}: phase diagram indicating the free energy, Eq.~\eqref{eq:free_energy_soft_wall}, of the black brane as a function of its temperature, Eq.~\eqref{eq:temperature_soft_wall} (black). The confined phase is normalized to $f = 0$ (gray). As the horizon at $y_h$ moves closer to the singularity, the temperature decreases to a minimum value $T_{\rm min}$, then increases. The minimum temperature $T_{\rm min}$ and the critical temperature $T_c$ are denoted by the red and blue dashed lines, respectively.
    \textbf{Right}: the effective potential in Eqs.~\eqref{eq:eff_potential_softwall_UV} and~\eqref{eq:eff_potential_softwall_IR}; the confined phase corresponds to the origin. For $T = T_c$ (blue), the minimum corresponding to the deconfined phase has $V = 0$. The deconfined phase is metastable for $T_{\rm min} < T < T_c$ (purple); at $T = T_{\rm min}$ (red) the minimum disappears, indicating there are no stable black brane solutions. }
    \label{fig:free_energy_potential}
\end{figure}

We can derive an analytical expression for the effective potential in Eq.~\eqref{eq:eff_potential_general}. In the UV regime, $y < y_i$, we write it in terms of the dimensionless field $\chi = k e^{-k y_h} / \pi T_c$, leading to
\begin{equation}\label{eq:eff_potential_softwall_UV}
    \frac{8}{\pi^2 N^2 T_c^4} V(y_h, T) = 3 \chi^4 - 4 \frac{T}{T_c} \chi^3 + 1 , \quad (y_h < y_i).
\end{equation}
In the IR regime $y > y_i$ it is easier to write the effective potential in terms of $\tau = k(y_s - y_h) (e^{-k y_i} k/ \pi T_c)^2$, yielding
\begin{equation}\label{eq:eff_potential_softwall_IR}
    \frac{8}{\pi^2 N^2 T_c^4} V(y_h, T) = -8\sqrt{2} \frac{T}{T_c} \tau^{3/2} - \frac{2}{2 \log 2 - 1} \log \left(1 - \sqrt{3(2 \log 2 - 1)} \tau \right) , \quad (y_h > y_i).
\end{equation}
In Eqs.~\eqref{eq:eff_potential_softwall_UV} and ~\eqref{eq:eff_potential_softwall_IR} we used the number of colors $N$ in the dual CFT, related to the 5D gravity parameters by the holographic relation $4\pi^2/N^2 = \kappa^2 k^3$.
We plot the effective potential in the right panel of Fig.~\ref{fig:free_energy_potential} for different values of $T$. The structure exhibited by the phase diagram is borne out by the effective potential. The black brane becomes metastable at $T = T_c$ and the metastable minimum disappears at $T = T_{\rm min}$, as expected.

\begin{figure}
    \centering
    \includegraphics[width=\textwidth]{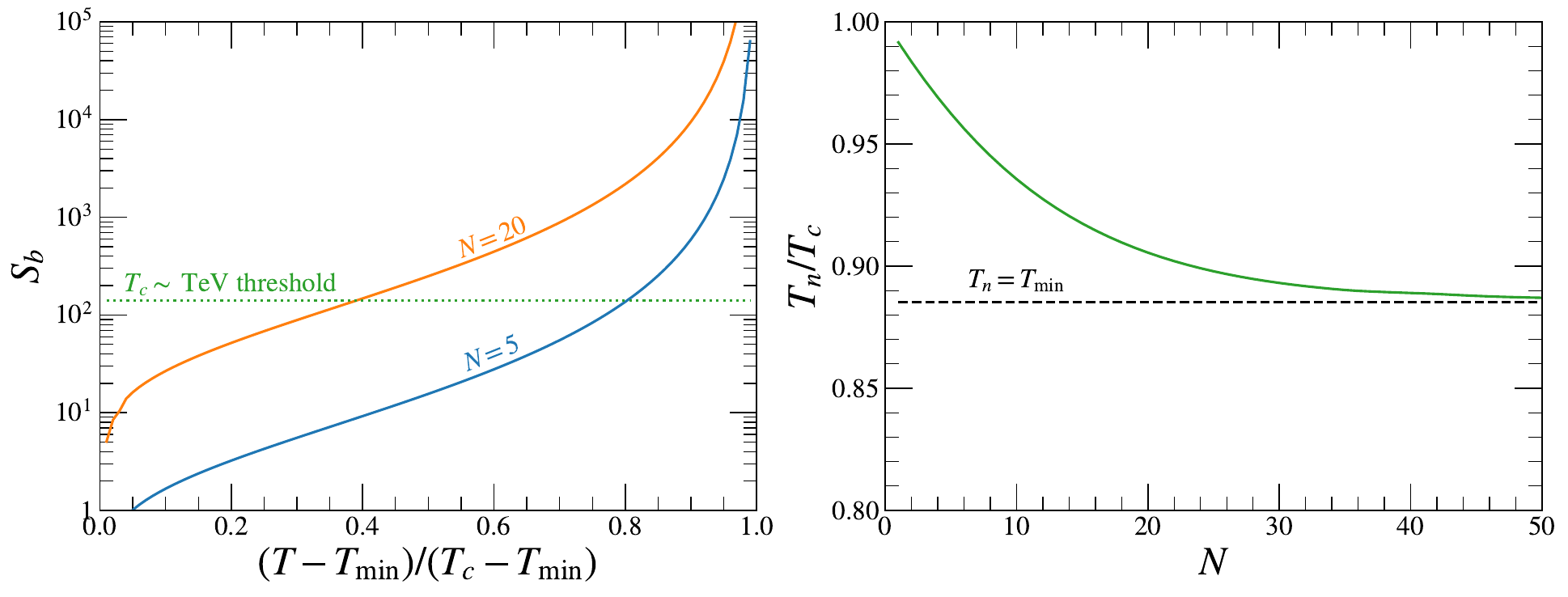}
    \caption{\textbf{Left}: the $O(3)$ bounce action $S_b = S_3 / T$ for the decay of the metastable black brane. The $x$-axis limits correspond to $T = T_{\rm min}$ and $T = T_c$. We show results for $N = 5$ (blue) and $N = 20$ (orange), which differ by a constant since $S_b \propto N^2$. The dashed green line is the threshold $S_b \approx 140$ for a TeV-scale PT to nucleate.
    \textbf{Right}: the amount of supercooling $T_n / T_c$ as a function of $N$. The nucleation temperature approaches the minimum temperature (dashed black line) as $N$ grows.}
    \label{fig:action_tn}
\end{figure}

It is also straightforward to derive the kinetic term using Eq.~\eqref{eq:kinetic_final}:
\begin{equation}\label{eq:kinetic_soft_wall}
    \frac{8\pi^2}{3 k^4 N^2} \mathcal{L}_{\rm kin} = \begin{cases}
        e^{-2 k y_h} (\nabla y_h)^2 & y < y_i \\
        e^{-2 k y_i} (\nabla y_h)^2 & y > y_i
    \end{cases}
\end{equation}
Armed with the effective action, we can calculate the $O(3)$-symmetric bounce action $S_3/T$~\cite{Coleman:1977py,Linde:1981zj}. We compute them numerically using the \texttt{FindBounce} package~\cite{Guada:2020xnz}; the results are presented in the left panel of Fig.~\ref{fig:action_tn}. We have checked that the numerical results agree with the thin-wall and thick-wall approximations as $T$ approaches $T_c$ and $T_{\rm min}$, respectively.

The phase transition completes when the bubble nucleation rate per unit volume, $\Gamma \sim T^4 \exp -S_b$, is larger than the Hubble parameter $H$. Assuming the universe is dominated by the energy density of the warped sector, we have $H^2 M_{\rm Pl}^2 \sim T_c^4$. This leads to a upper bound on the bounce action for the phase transition to complete of~\cite{Agashe:2019lhy,vonHarling:2017yew}
\begin{equation}\label{eq:nucleation_condition}
    S_b \lesssim 4 \log \frac{M_{\rm Pl}}{T_c} .
\end{equation}
For a TeV-scale PT, characteristic of applications to the hierarchy problem, we find a threshold of $S_b \lesssim 140$. Using this we estimate the nucleation temperature $T_n$ at which the PT completes in the right panel of Fig.~\ref{fig:action_tn}. Since the bounce action grows as $N^2$, as $N$ increases the nucleation temperature approaches $T_{\rm min}$. Furthermore, it is clear that the phase transition completes without much supercooling.

\subsection{Gravitational wave signals}\label{sec:GW}

First-order PTs can source a stochastic gravitational wave background through bubble collisions, sound waves, and turbulence in the plasma.
The important parameters characterizing the GW signal are the inverse duration of the PT $\beta$, the strength $\alpha_{\rm PT}$, and the bubble wall velocity $v_w$.

\begin{figure}
    \centering
    \includegraphics[width=\textwidth]{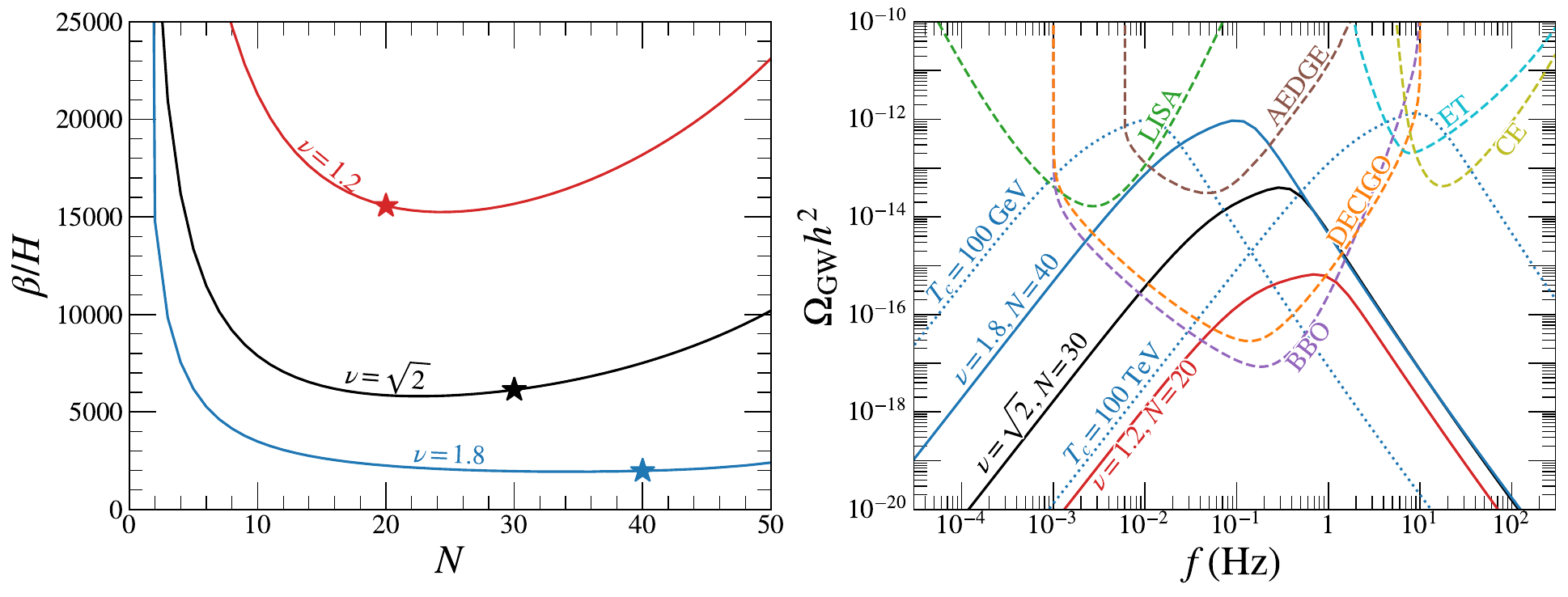}
    \caption{\textbf{Left}: the inverse PT duration $\beta/H$ computed with Eq.~\eqref{eq:beta_over_H}, taking $T_c = 1$~TeV. We show results for $\nu = 1.2$ (red), $\nu = \sqrt{2}$ (black), and $\nu = 1.8$ (blue) as a function of $N$. The stars indicate three benchmark points.
    \textbf{Right}: the stochastic GW background from sound waves and bubble collisions for these three benchmark points, taking $T_c = 1$~TeV (solid lines). For the blue benchmark we show the effect of taking $T_c = 100$~GeV or $100$~TeV (dotted lines). We include projected sensitivities, adapted from~\cite{Schmitz:2020syl}, for LISA (green)~\cite{LISA:2017pwj,Baker:2019nia}, BBO (purple)~\cite{Crowder:2005nr,Corbin:2005ny,Harry:2006fi}, DECIGO (orange)~\cite{Seto:2001qf,Kawamura:2011zz,Yagi:2011wg,Isoyama:2018rjb}, AEDGE (brown)~\cite{AEDGE:2019nxb,Badurina:2019hst}, Cosmic Explorer (CE, olive)~\cite{LIGOScientific:2016wof,Reitze:2019iox}, and  Einstein Telescope (ET, cyan)~\cite{Punturo:2010zz,Hild:2010id,Sathyaprakash:2012jk,ET:2019dnz}.}
    \label{fig:betaOverH_GW}
\end{figure}

The inverse duration can be derived from the bounce action as~\cite{Caprini:2024hue}
\begin{equation}\label{eq:beta_over_H}
    \frac{\beta}{H_{\rm PT}} = T \frac{dS_b}{dT} \Big |_{T = T_n} .
\end{equation}
In the left panel of Fig.~\ref{fig:betaOverH_GW} we plot $\beta/H$ using Eq.~\eqref{eq:beta_over_H}, again assuming a TeV-scale PT.
We present results for multiple values of the superpotential growth rate, $\nu = 1.2, \sqrt{2}, 1.8$.
Since the PT is not strongly supercooled, $\beta/H$ is larger as compared to a supercooled conformal PT. We find $\beta/H \sim 10^{3\text{--}4}$ is typical; in contrast, $\beta/H \sim 10$ is typical in Goldberger--Wise stabilization. As $\nu$ increases, more supercooling is allowed, so $\beta/H$ decreases. We mark three benchmark points with stars: $\nu = 1.2, N = 20$, $\nu = \sqrt{2}, N = 30$, and $\nu = 1.8, N = 40$.

The strength $\alpha_{\rm PT}$ is defined as the ratio of the latent heat released in the PT to the energy density of the radiation bath. The latent heat is $f(T_n)$, while the radiation energy density is of order $\pi^2 N^2 T_n^4/30$, so
\begin{equation}\label{eq:pt_strength}
    \alpha_{\rm PT} = \frac{30 f(T_n)}{\pi^2 N^2 T_n^4} .
\end{equation}
Given that $f \sim N^2 T_c^4$, we expect $\alpha_{\rm PT}$ to be order-one. Indeed, for the benchmark points in Fig.~\ref{fig:betaOverH_GW}, this equation gives $\alpha_{\rm PT} \approx 0.37, 1.65, 6.89$, in order of increasing $\nu$ and $N$.

The bubble wall velocity is notoriously difficult to calculate. Highly relativistic wall velocities are typical for the values of $\alpha \gtrsim 0.1$ relevant to us~\cite{Laurent:2022jrs}. There has also been recent progress in studying the bubble wall velocity in holographic scenarios~\cite{Bigazzi:2021ucw,Bea:2021zsu}. The results of~\cite{Bigazzi:2021ucw} also suggest a highly relativistic wall for the values of $\alpha$ we consider. For these reasons, we shall assume $v_w \approx 1$.

As a consequence of the relatively fast PT, the expected GW signals are weaker.
We plot the GW signal for our three benchmark points in the right panel of Fig.~\ref{fig:betaOverH_GW}, alongside projected sensitivities for future space-based and ground-based detectors~\cite{Schmitz:2020syl}. For the most optimistic benchmark ($\nu = 1.8, N = 40$) we also show signal curves for a lower-scale, $100$~GeV PT and a higher-scale, $100$~TeV PT. We compute the spectrum using the broken power law templates adopted by the LISA Cosmology Working Group~\cite{Caprini:2024hue}. We include contributions from bubble collisions and from sound waves. However, if we only include bubble collisions or only include sound waves, this does not affect our conclusions regarding detection prospects. As we cannot estimate the fraction of kinetic energy that is converted into turbulence, we conservatively neglect this contribution.

The space-based detectors BBO~\cite{Crowder:2005nr,Corbin:2005ny,Harry:2006fi} and DECIGO~\cite{Seto:2001qf,Kawamura:2011zz,Yagi:2011wg,Isoyama:2018rjb} could probe all three of our benchmark points.
AEDGE~\cite{AEDGE:2019nxb,Badurina:2019hst} could detect the most optimistic benchmark but not the other two.
Considering lower or higher $T_c$ can shift the signal into the realm of what can be detected at LISA or terrestrial interferometers, respectively. Our optimistic benchmark with $T_c = 100$~GeV is within reach of LISA~\cite{LISA:2017pwj,Baker:2019nia}, while a $T_c = 100$~TeV transition could be probed at the terrestrial Einstein Telescope~\cite{Punturo:2010zz,Hild:2010id,Sathyaprakash:2012jk,ET:2019dnz} and Cosmic Explorer~\cite{LIGOScientific:2016wof,Reitze:2019iox}. It is encouraging that even in this scenario with a relatively weak PT, there is still a possibility of discovering the stochastic GW background at next-generation experiments.

\section{On the edge of confinement}\label{sec:linear_dilaton}

Recall that for a superpotential that grows as $W \sim \exp \nu \kappa \phi / \sqrt{3}$, one finds a confining geometry for $1 \leq \nu < 2$. In this section we discuss the conformal PT in the edge case $\nu = 1$. This is qualitatively different from $1 < \nu < 2$ because the subleading behavior of the metric near the singularity is important.

We consider a superpotential that grows as
\begin{equation}
    W \sim \phi^{p/2} e^{\kappa \phi / \sqrt{3}} .
\end{equation}
As shown in~\cite{Gursoy:2007cb,Gursoy:2007er}, near the singularity the warp factor behaves as 
\begin{equation}\label{eq:IR_asymptotics_subleading}
    A \sim - \log k (y_s - y) - p \log \left( - \log k (y_s - y) \right) .
\end{equation}
The first term is the leading result in Eq.~\eqref{eq:IR_asymptotics}, while the second term is the subleading behavior.

There are a couple of specific cases that are worth mentioning explicitly. The case $p = 0$ corresponds to a linear dilaton geometry. To see this, observe that $A \sim - \log( 1 - y/y_s)$ near the singularity, c.f. Eq.~\eqref{eq:IR_asymptotics_subleading}. We can rewrite the metric, Eq.~\eqref{eq:metric_ansatz_cold} in terms of the conformal coordinate $z$, where it takes the form
\begin{equation}
    ds^2 = e^{-2A(z)} \left( \eta_{\mu\nu} dx^\mu dx^\nu - dz^2 \right) .
\end{equation}
The conformal coordinate can be related to the $y$ coordinate as $dz/dy = e^A$, from which it follows that $A(z) \sim z$ --- a linear dilaton. One can further show that such a geometry leads to a gapped continuum spectrum for KK modes, which is of phenomenological interest~\cite{Falkowski:2008fz,Falkowski:2008yr,Cabrer:2009we,Bellazzini:2015cgj,Csaki:2018kxb,Megias:2019vdb,Megias:2021mgj,Csaki:2021gfm,Csaki:2021xpy,Csaki:2022lnq,Fichet:2022xol,Fichet:2023dju,Fichet:2023xbu,Ferrante:2023fpx,Ferrante:2025ofe}.
In a similar fashion, one can show that the case $p = 1/2$ corresponds to $A(z) \sim z^2$. Such a geometry is interesting from the perspective of AdS/QCD, as it leads to linear Regge trajectories~\cite{Karch:2006pv}.

As in Section~\ref{sec:main_result}, we consider a piecewise ansatz for the warp factor:
\begin{equation}
    A'(y) = \begin{cases}
        k  & y < y_i \\
        \frac{1}{y_s - y} \left( 1 +  \frac{ p }{ \log k (y_s - y)}  \right)  & y > y_i.
    \end{cases}
\end{equation}
The matching point $\psi_i  = k(y_s - y_i)$ is determined by the solution to $\psi_i = 1 + p / \log \psi_i$. This does not admit an analytical solution (except when $p = 0$). By integrating the above equation we can find the warp factor. Recall also from Eq.~\eqref{eq:einstein_equations_hot} that the blackening factor is determined by $b'' = 4 A' b'$, together with the boundary conditions that $b \rightarrow 1$ as $y \rightarrow -\infty$ and $b(y_h) = 0$. Using this we can numerically compute the temperature, Eq.~\eqref{eq:temperature}, and the free energy, Eq.~\eqref{eq:free_energy}.

\begin{figure}
    \centering
    \includegraphics[width=\textwidth]{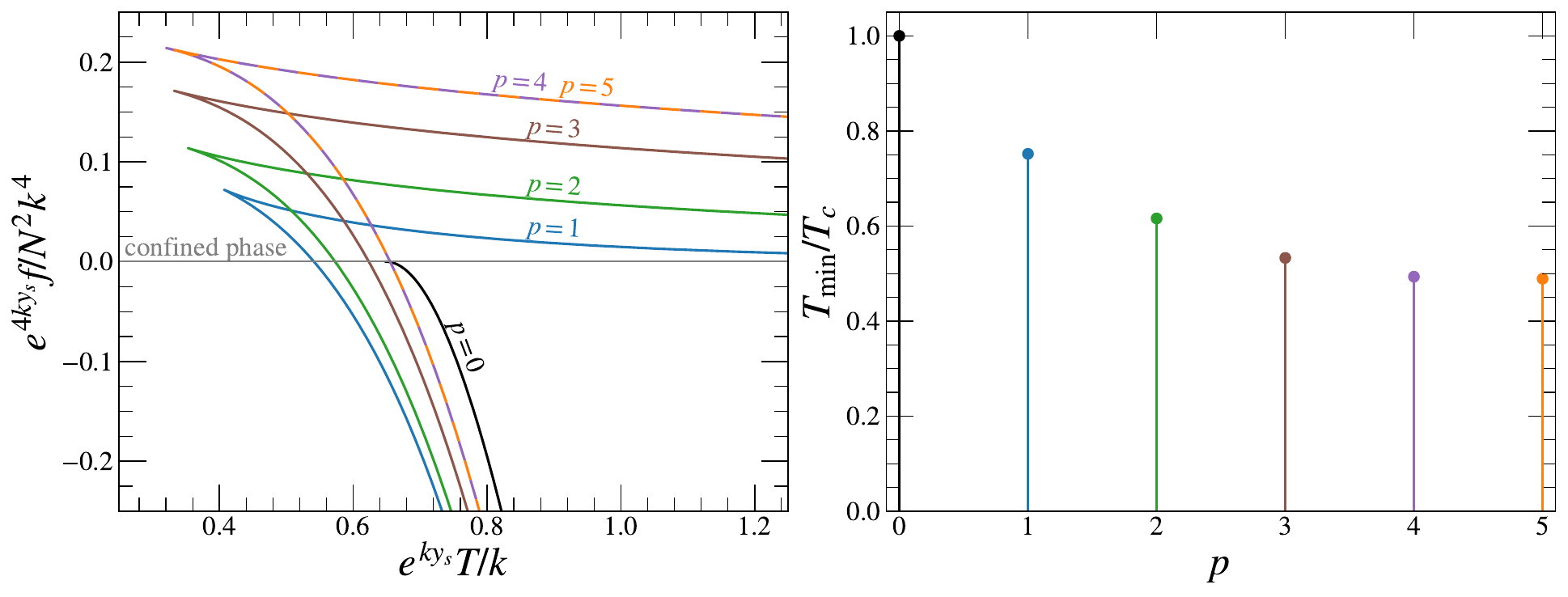}
    \caption{\textbf{Left}: phase diagram for confinement with a superpotential $W \sim \phi^{p/2} \exp \kappa \phi / \sqrt{3}$. The confined phase is normalized to $f = 0$ (gray). The curves for $p = 4$ and $p = 5$ are indistinguishable.
    \textbf{Right}: the ratio of the minimum temperature to the critical temperature $T_{\rm min} / T_c$ as a function of $p$. }
    \label{fig:edge_confinement}
\end{figure}

In Fig.~\ref{fig:edge_confinement} we plot phase diagrams for integer values of $p$ from $0$ to $5$. With the exception of $p = 0$, they all exhibit the same shark-fin shape, characterized by two branches of black brane solutions that meet at the minimum temperature $T_{\rm min}$. As $p$ increases the phase diagram appears to approach a limiting form. We numerically compute the critical temperature $T_c$ by solving for $f = 0$ and plot the maximum possible supercooling $T_{\rm min}/T_c$ in Fig.~\ref{fig:edge_confinement} as well. As $p$ increases the ratio $T_{\rm min} / T_c$ approaches a limiting value of about $1/2$. It would be interesting to demonstrate this behavior analytically. In any case, it is clear that there is a qualitative similarity with the soft walls studied in Section~\ref{sec:main_result} for $p \neq 0$.

The linear dilaton geometry, $p = 0$, is different. From Fig.~\ref{fig:edge_confinement} we observe a single branch of black branes, which ends at $T_{\rm min}$. The free energy is negative except at $T_{\rm min}$, where it vanishes. This means that the critical temperature is the minimum temperature and no supercooling is possible. This is a second order phase transition.

To better understand the linear dilaton PT, we can derive analytical expressions for the temperature and free energy (which we could not do for $p \neq 0$). The metric is just given by Eq.~\eqref{eq:warp_factor_deriv} with $\nu = 1$. For the temperature we find
\begin{equation}
    \pi \frac{T_h}{k} = \begin{cases}
        e^{-k y_h} & y < y_i \\
        e^{-k y_i} \frac{3}{4 -\psi^3} & y > y_i
    \end{cases} ,
\end{equation}
where $y_i = y_s - 1/k$ and $\psi = k(y_s - y_h)$. As $y_h$ increases towards $y_s$, the temperature decreases monotonically, approaching the minimum value $T_{\rm min} = 3k e^{-k y_i} / 4\pi$.

For the free energy we find
\begin{equation}
    2\frac{\kappa^2}{k} f(y_h) = \begin{cases}
        -e^{-4 k y_h} + 3 (4 \log 4/3 - 1) e^{-4 k y_i} & y < y_i \\
        12 e^{-4 y_i} \left(1 - \frac{4}{4-\psi^3} + \log \frac{4}{4-\psi^3} \right) & y > y_i .
    \end{cases}
\end{equation}
Recall that we choose the overall constant so $f(y_s) = 0$.
Crucially, we see that $f(y_h) < 0$ for all $y_h < y_s$. This indicates that whenever the black brane solution exists, it is thermodynamically favored over the confined phase. The black brane solution does not exist below $T_{\rm min}$, corresponding to $y_h \rightarrow y_s$. We conclude that the confinement PT is second order and takes place at $T = T_{\rm min}$, which is in agreement with~\cite{Gursoy:2007cb,Gursoy:2008za,Gursoy:2007er}.

\begin{figure}
    \centering
    \includegraphics[width=\textwidth]{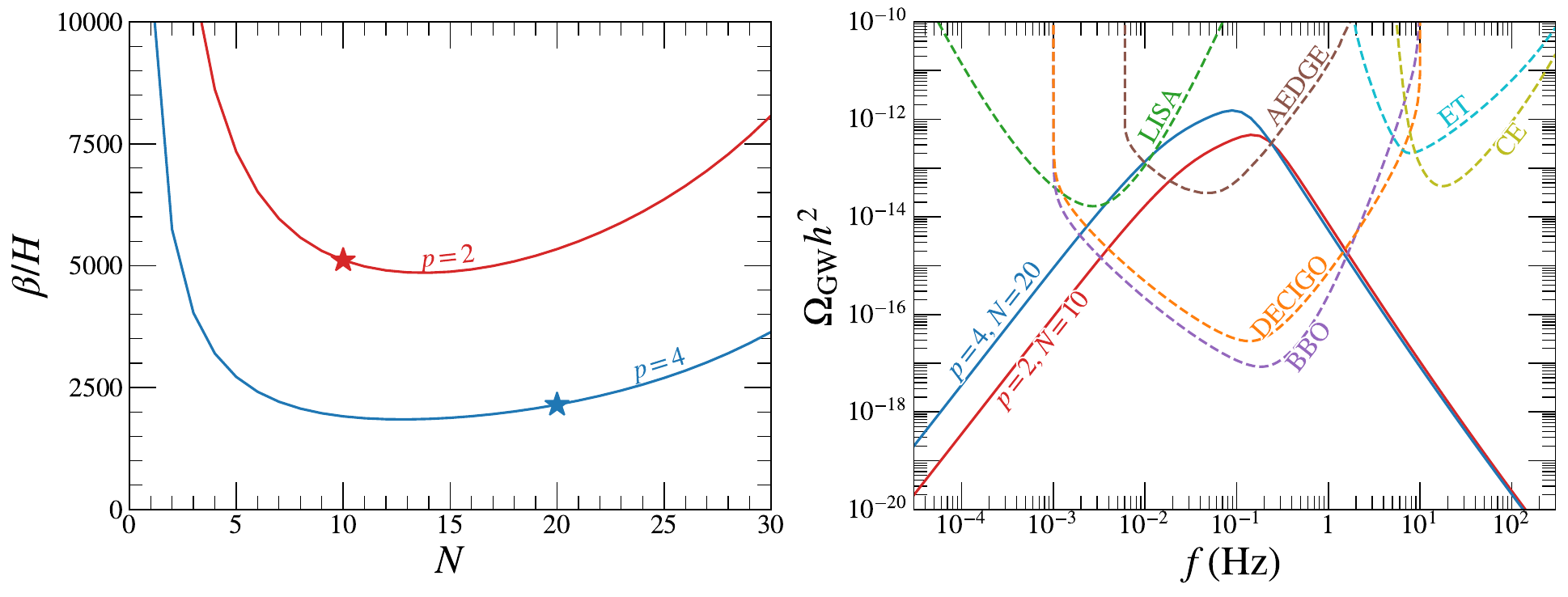}
    \caption{\textbf{Left}: the inverse PT duration $\beta/H$ for a superpotential $W \sim \phi^{p/2} \exp \kappa \phi / \sqrt{3}$. We show results for $p = 2$ (red) and $p = 4$ (blue) as a function of $N$. The stars indicate two benchmark points.
    \textbf{Right}: the stochastic GW background from sound waves and bubble collisions for these two benchmark points. The experimental projections are the same as Fig.~\ref{fig:betaOverH_GW}.}
    \label{fig:edge_GW}
\end{figure}

For $p > 0$ the transition is first order and we expect a stochastic GW background. Although we regard the geometries studied in Section~\ref{sec:main_result} as better motivated, we calculate the GW signal anyway for completeness. We numerically compute the effective action with Eqs.~\eqref{eq:eff_potential_general} and~\eqref{eq:kinetic_final}. Assuming a TeV-scale PT, we find the nucleation temperature with Eq.~\eqref{eq:nucleation_condition}. We calculate the inverse PT duration $\beta/H$ using Eq.~\eqref{eq:beta_over_H}, which we present in the left panel of Fig.~\ref{fig:edge_GW} for $p = 2$ and $p = 4$. Like in the geometries with $\nu > 1$ (see Fig.~\ref{fig:betaOverH_GW}), $\beta/H \sim 10^3$ is typical.

We select two benchmark points ($p = 2$, $N = 10$ and $p = 4, N = 20$); we find $\alpha_{\rm PT} > 10$ for both using Eq.~\eqref{eq:pt_strength}. In the right panel of Fig.~\ref{fig:edge_GW} we show the GW background resulting from sound waves and bubble collisions. One of our benchmarks can just barely be probed by LISA, while AEDGE, BBO, and DECIGO would be sensitive to both benchmarks.

\section{Conclusions}\label{sec:conclusions}
Warped extra dimensions constitute a class of well-motivated new physics models that can enjoy a strong first-order PT. It is crucial to understand the dynamics of the PT and the resulting stochastic GW background to assess the prospects for discovery at future GW detectors.
In this work we studied the confinement PT in a warped extra dimension cut off by a curvature singularity. This was motivated by the dearth of such studies in models with soft walls as opposed to IR branes; by recent work suggesting that the dynamics are remarkably different from standard RS lore~\cite{Mishra:2024ehr}; and by the occurrence of qualitatively similar phenomena in top-down constructions~\cite{Buchel:2021yay,Buchel:2018bzp,Buchel:2024phy}.

We worked within an approximation where the bounce was parametrized by a single field corresponding to the black brane horizon location. Ideally, we would like to do a full 5D computation of the bounce by solving the Euclidean-time Einstein equations. Although it is challenging numerically, similar computations have been performed in~\cite{Aharony:2005bm}, and more recently codes for studying holographic bubbles have been developed~\cite{Bea:2022mfb}. This would serve as a useful check of the results obtained here.

We focused on a confining superpotential growing asymptotically as $\phi^n \exp \nu \kappa \phi/\sqrt{3}$, with $1 < \nu < 2$. We considered an ansatz for the warp factor by stitching together the UV AdS solution and the IR near-singularity solution. This is the simplest possibility that still allowed us to capture the essence of the PT in soft-wall models. We confirmed that the black brane solutions exhibited a minimum temperature, indicating that not much supercooling is possible. We further verified this by studying the nucleation temperature assuming a TeV-scale PT. Despite the weaker PT, the GW signals would still be accessible to future space-based interferometers. In optimistic scenarios there is a prospect for discovery at AEDGE, while other parts of parameter space would require BBO or DECIGO to probe. A lower critical temperature of $100$~GeV could be accessible to LISA, while a $100$~TeV-scale PT could be probed by future terrestrial interferometers.

We also studied the thermodynamics in the edge case $\nu = 1$. This encompasses some geometries that are phenomenologically interesting, including the linear dilaton and one that gives rise to linear confinement. We found the phase structure is similar to the soft walls with $1 < \nu < 2$, except for the linear dilaton geometry. The expected GW signals are similar as well, with $\beta/H \sim 10^3$ and our benchmarks accessible to AEDGE, BBO, and DECIGO. In the linear dilaton case, the phase transition is second order.

As a future direction, it would be useful to study the case where the universe starts out on the unstable black hole branch. This would be a unique test of the phase diagram. It was shown in~\cite{Bea:2021zol} that the decay to the stable branch can source GWs. However, a numerical simulation would likely be necessary to understand the dynamics, which are more complicated than the standard first-order PT picture.

\begin{acknowledgments}
AI is supported by a Mafalda and Reinhard Oehme Postdoctoral Research Fellowship from the Enrico Fermi Institute at the University of Chicago.
LTW is supported by the Department of Energy grant DE-SC0013642.

\noindent\textit{Disclaimer}: all numerical calculations were performed by AI\footnote{Ameen Ismail}, not by AI\footnote{artificial intelligence}.

\end{acknowledgments}

\bibliographystyle{JHEP}
\bibliography{references}

@article{Randall:1999ee,
    author = "Randall, Lisa and Sundrum, Raman",
    title = "{A Large mass hierarchy from a small extra dimension}",
    eprint = "hep-ph/9905221",
    archivePrefix = "arXiv",
    reportNumber = "MIT-CTP-2860, PUPT-1860, BUHEP-99-9",
    doi = "10.1103/PhysRevLett.83.3370",
    journal = "Phys. Rev. Lett.",
    volume = "83",
    pages = "3370--3373",
    year = "1999"
}

@article{Maldacena:1997re,
    author = "Maldacena, Juan Martin",
    title = "{The Large $N$ limit of superconformal field theories and supergravity}",
    eprint = "hep-th/9711200",
    archivePrefix = "arXiv",
    reportNumber = "HUTP-97-A097, HUTP-98-A097",
    doi = "10.4310/ATMP.1998.v2.n2.a1",
    journal = "Adv. Theor. Math. Phys.",
    volume = "2",
    pages = "231--252",
    year = "1998"
}

@article{Witten:1998qj,
    author = "Witten, Edward",
    title = "{Anti de Sitter space and holography}",
    eprint = "hep-th/9802150",
    archivePrefix = "arXiv",
    reportNumber = "IASSNS-HEP-98-15",
    doi = "10.4310/ATMP.1998.v2.n2.a2",
    journal = "Adv. Theor. Math. Phys.",
    volume = "2",
    pages = "253--291",
    year = "1998"
}

@article{Gubser:1998bc,
    author = "Gubser, S. S. and Klebanov, Igor R. and Polyakov, Alexander M.",
    title = "{Gauge theory correlators from noncritical string theory}",
    eprint = "hep-th/9802109",
    archivePrefix = "arXiv",
    reportNumber = "PUPT-1767",
    doi = "10.1016/S0370-2693(98)00377-3",
    journal = "Phys. Lett. B",
    volume = "428",
    pages = "105--114",
    year = "1998"
}

@article{Arkani-Hamed:2000ijo,
    author = "Arkani-Hamed, Nima and Porrati, Massimo and Randall, Lisa",
    title = "{Holography and phenomenology}",
    eprint = "hep-th/0012148",
    archivePrefix = "arXiv",
    reportNumber = "LBL-46886, UCB-PTH-00-32, NYU-TH-00-09-01",
    doi = "10.1088/1126-6708/2001/08/017",
    journal = "JHEP",
    volume = "08",
    pages = "017",
    year = "2001"
}

@article{Rattazzi:2000hs,
    author = "Rattazzi, R. and Zaffaroni, A.",
    title = "{Comments on the holographic picture of the Randall-Sundrum model}",
    eprint = "hep-th/0012248",
    archivePrefix = "arXiv",
    reportNumber = "SNS-PH-00-21, BICOCCA-FT-00-25",
    doi = "10.1088/1126-6708/2001/04/021",
    journal = "JHEP",
    volume = "04",
    pages = "021",
    year = "2001"
}

@article{Goldberger:1999uk,
    author = "Goldberger, Walter D. and Wise, Mark B.",
    title = "{Modulus stabilization with bulk fields}",
    eprint = "hep-ph/9907447",
    archivePrefix = "arXiv",
    reportNumber = "CALT-68-2232",
    doi = "10.1103/PhysRevLett.83.4922",
    journal = "Phys. Rev. Lett.",
    volume = "83",
    pages = "4922--4925",
    year = "1999"
}

@article{Creminelli:2001th,
    author = "Creminelli, Paolo and Nicolis, Alberto and Rattazzi, Riccardo",
    title = "{Holography and the electroweak phase transition}",
    eprint = "hep-th/0107141",
    archivePrefix = "arXiv",
    reportNumber = "CERN-TH-2001-189",
    doi = "10.1088/1126-6708/2002/03/051",
    journal = "JHEP",
    volume = "03",
    pages = "051",
    year = "2002"
}

@article{Randall:2006py,
    author = "Randall, Lisa and Servant, Geraldine",
    title = "{Gravitational waves from warped spacetime}",
    eprint = "hep-ph/0607158",
    archivePrefix = "arXiv",
    reportNumber = "CERN-PH-TH-2006-133",
    doi = "10.1088/1126-6708/2007/05/054",
    journal = "JHEP",
    volume = "05",
    pages = "054",
    year = "2007"
}

@article{Konstandin:2010cd,
    author = "Konstandin, Thomas and Nardini, Germano and Quiros, Mariano",
    title = "{Gravitational Backreaction Effects on the Holographic Phase Transition}",
    eprint = "1007.1468",
    archivePrefix = "arXiv",
    primaryClass = "hep-ph",
    reportNumber = "CERN-PH-TH-2010-151, UAB-FT-683, ULB-TH-10-19",
    doi = "10.1103/PhysRevD.82.083513",
    journal = "Phys. Rev. D",
    volume = "82",
    pages = "083513",
    year = "2010"
}

@article{Konstandin:2011dr,
    author = "Konstandin, Thomas and Servant, Geraldine",
    title = "{Cosmological Consequences of Nearly Conformal Dynamics at the TeV scale}",
    eprint = "1104.4791",
    archivePrefix = "arXiv",
    primaryClass = "hep-ph",
    doi = "10.1088/1475-7516/2011/12/009",
    journal = "JCAP",
    volume = "12",
    pages = "009",
    year = "2011"
}

@article{vonHarling:2017yew,
    author = "von Harling, Benedict and Servant, Geraldine",
    title = "{QCD-induced Electroweak Phase Transition}",
    eprint = "1711.11554",
    archivePrefix = "arXiv",
    primaryClass = "hep-ph",
    reportNumber = "DESY-17-056",
    doi = "10.1007/JHEP01(2018)159",
    journal = "JHEP",
    volume = "01",
    pages = "159",
    year = "2018"
}

@article{Bruggisser:2018mus,
    author = "Bruggisser, Sebastian and Von Harling, Benedict and Matsedonskyi, Oleksii and Servant, G{\'e}raldine",
    title = "{Baryon Asymmetry from a Composite Higgs Boson}",
    eprint = "1803.08546",
    archivePrefix = "arXiv",
    primaryClass = "hep-ph",
    reportNumber = "DESY-18-029",
    doi = "10.1103/PhysRevLett.121.131801",
    journal = "Phys. Rev. Lett.",
    volume = "121",
    number = "13",
    pages = "131801",
    year = "2018"
}

@article{Bruggisser:2018mrt,
    author = "Bruggisser, Sebastian and Von Harling, Benedict and Matsedonskyi, Oleksii and Servant, G{\'e}raldine",
    title = "{Electroweak Phase Transition and Baryogenesis in Composite Higgs Models}",
    eprint = "1804.07314",
    archivePrefix = "arXiv",
    primaryClass = "hep-ph",
    reportNumber = "DESY-17-229",
    doi = "10.1007/JHEP12(2018)099",
    journal = "JHEP",
    volume = "12",
    pages = "099",
    year = "2018"
}

@article{Baratella:2018pxi,
    author = "Baratella, Pietro and Pomarol, Alex and Rompineve, Fabrizio",
    title = "{The Supercooled Universe}",
    eprint = "1812.06996",
    archivePrefix = "arXiv",
    primaryClass = "hep-ph",
    doi = "10.1007/JHEP03(2019)100",
    journal = "JHEP",
    volume = "03",
    pages = "100",
    year = "2019"
}

@article{Agashe:2019lhy,
    author = "Agashe, Kaustubh and Du, Peizhi and Ekhterachian, Majid and Kumar, Soubhik and Sundrum, Raman",
    title = "{Cosmological Phase Transition of Spontaneous Confinement}",
    eprint = "1910.06238",
    archivePrefix = "arXiv",
    primaryClass = "hep-ph",
    reportNumber = "UMD-PP-019-05, YITP-SB-19-32",
    doi = "10.1007/JHEP05(2020)086",
    journal = "JHEP",
    volume = "05",
    pages = "086",
    year = "2020"
}

@article{Agashe:2020lfz,
    author = "Agashe, Kaustubh and Du, Peizhi and Ekhterachian, Majid and Kumar, Soubhik and Sundrum, Raman",
    title = "{Phase Transitions from the Fifth Dimension}",
    eprint = "2010.04083",
    archivePrefix = "arXiv",
    primaryClass = "hep-th",
    reportNumber = "UMD-PP-020-5, YITP-SB-2020-29",
    doi = "10.1007/JHEP02(2021)051",
    journal = "JHEP",
    volume = "02",
    pages = "051",
    year = "2021"
}

@article{Agrawal:2021alq,
    author = "Agrawal, Prateek and Nee, Michael",
    title = "{Avoided deconfinement in Randall-Sundrum models}",
    eprint = "2103.05646",
    archivePrefix = "arXiv",
    primaryClass = "hep-ph",
    doi = "10.1007/JHEP10(2021)105",
    journal = "JHEP",
    volume = "10",
    pages = "105",
    year = "2021"
}

@article{Baldes:2021aph,
    author = "Baldes, Iason and Gouttenoire, Yann and Sala, Filippo and Servant, G{\'e}raldine",
    title = "{Supercool composite Dark Matter beyond 100 TeV}",
    eprint = "2110.13926",
    archivePrefix = "arXiv",
    primaryClass = "hep-ph",
    reportNumber = "ULB-TH/21-17; DESY 21-172, ULB-TH/21-17, DESY 21-172",
    doi = "10.1007/JHEP07(2022)084",
    journal = "JHEP",
    volume = "07",
    pages = "084",
    year = "2022"
}

@article{Bruggisser:2022rdm,
    author = "Bruggisser, Sebastian and von Harling, Benedict and Matsedonskyi, Oleksii and Servant, Geraldine",
    title = "{Status of electroweak baryogenesis in minimal composite Higgs}",
    eprint = "2212.11953",
    archivePrefix = "arXiv",
    primaryClass = "hep-ph",
    reportNumber = "DESY-22-209",
    doi = "10.1007/JHEP08(2023)012",
    journal = "JHEP",
    volume = "08",
    pages = "012",
    year = "2023"
}

@article{Csaki:2023pwy,
    author = "Cs{\'a}ki, Csaba and Geller, Michael and Heller-Algazi, Zamir and Ismail, Ameen",
    title = "{Relevant dilaton stabilization}",
    eprint = "2301.10247",
    archivePrefix = "arXiv",
    primaryClass = "hep-ph",
    doi = "10.1007/JHEP06(2023)202",
    journal = "JHEP",
    volume = "06",
    pages = "202",
    year = "2023"
}

@article{Eroncel:2023uqf,
    author = {Er{\"o}ncel, Cem and Hubisz, Jay and Lee, Seung. J. and Rigo, Gabriele and Sambasivam, Bharath},
    title = "{New horizons in the holographic conformal phase transition}",
    eprint = "2305.03773",
    archivePrefix = "arXiv",
    primaryClass = "hep-ph",
    doi = "10.1140/epjc/s10052-024-13125-6",
    journal = "Eur. Phys. J. C",
    volume = "84",
    number = "8",
    pages = "794",
    year = "2024"
}

@article{Megias:2023kiy,
    author = "Megias, Eugenio and Nardini, Germano and Quiros, Mariano",
    title = "{Pulsar timing array stochastic background from light Kaluza-Klein resonances}",
    eprint = "2306.17071",
    archivePrefix = "arXiv",
    primaryClass = "hep-ph",
    doi = "10.1103/PhysRevD.108.095017",
    journal = "Phys. Rev. D",
    volume = "108",
    number = "9",
    pages = "095017",
    year = "2023"
}

@article{Ferrante:2023bcz,
    author = "Ferrante, Steven and Ismail, Ameen and Lee, Seung J. and Lee, Yunha",
    title = "{Forbidden conformal dark matter at a GeV}",
    eprint = "2308.16219",
    archivePrefix = "arXiv",
    primaryClass = "hep-ph",
    doi = "10.1007/JHEP11(2023)186",
    journal = "JHEP",
    volume = "11",
    pages = "186",
    year = "2023"
}

@article{Mishra:2023kiu,
    author = "Mishra, Rashmish K. and Randall, Lisa",
    title = "{Consequences of a stabilizing field{\textquoteright}s self-interactions for RS cosmology}",
    eprint = "2309.10090",
    archivePrefix = "arXiv",
    primaryClass = "hep-ph",
    doi = "10.1007/JHEP12(2023)036",
    journal = "JHEP",
    volume = "12",
    pages = "036",
    year = "2023"
}

@article{Luo:2025alo,
    author = "Luo, Lillian and Perelstein, Maxim",
    title = "{Conformal Freeze-in dark matter: 5D dual and phase transition}",
    eprint = "2502.06965",
    archivePrefix = "arXiv",
    primaryClass = "hep-ph",
    doi = "10.1007/JHEP06(2025)154",
    journal = "JHEP",
    volume = "06",
    pages = "154",
    year = "2025"
}

@article{Agrawal:2025wvf,
    author = "Agrawal, Prateek and Kane, Gaurang Ramakant and Loladze, Vazha and Reig, Mario",
    title = "{Supercooled confinement}",
    eprint = "2504.00199",
    archivePrefix = "arXiv",
    primaryClass = "hep-ph",
    doi = "10.1007/JHEP10(2025)066",
    journal = "JHEP",
    volume = "10",
    pages = "066",
    year = "2025"
}

@article{Mishra:2026lvq,
    author = "Mishra, Rashmish K.",
    title = "{Confinement in Holographic Theories at Finite Theta}",    eprint = "2603.24732",
    archivePrefix = "arXiv",
    primaryClass = "hep-th",
    month = "3",
    year = "2026"
}

@article{Mishra:2024ehr,
    author = "Mishra, Rashmish K. and Randall, Lisa",
    title = "{Phase transition to RS: cool, not supercool}",
    eprint = "2401.09633",
    archivePrefix = "arXiv",
    primaryClass = "hep-ph",
    doi = "10.1007/JHEP06(2024)099",
    journal = "JHEP",
    volume = "06",
    pages = "099",
    year = "2024"
}

@article{Gursoy:2008za,
    author = "Gursoy, U. and Kiritsis, E. and Mazzanti, L. and Nitti, F.",
    title = "{Holography and Thermodynamics of 5D Dilaton-gravity}",
    eprint = "0812.0792",
    archivePrefix = "arXiv",
    primaryClass = "hep-th",
    reportNumber = "CPHT-RR088-1108, SPIN-08-57, ITP-UU-08-74",
    doi = "10.1088/1126-6708/2009/05/033",
    journal = "JHEP",
    volume = "05",
    pages = "033",
    year = "2009"
}

@article{Bigazzi:2020avc,
    author = "Bigazzi, Francesco and Caddeo, Alessio and Cotrone, Aldo L. and Paredes, Angel",
    title = "{Dark Holograms and Gravitational Waves}",
    eprint = "2011.08757",
    archivePrefix = "arXiv",
    primaryClass = "hep-ph",
    doi = "10.1007/JHEP04(2021)094",
    journal = "JHEP",
    volume = "04",
    pages = "094",
    year = "2021"
}

@article{Morgante:2022zvc,
    author = "Morgante, Enrico and Ramberg, Nicklas and Schwaller, Pedro",
    title = "{Gravitational waves from dark SU(3) Yang-Mills theory}",
    eprint = "2210.11821",
    archivePrefix = "arXiv",
    primaryClass = "hep-ph",
    doi = "10.1103/PhysRevD.107.036010",
    journal = "Phys. Rev. D",
    volume = "107",
    number = "3",
    pages = "036010",
    year = "2023"
}

@article{Ares:2020lbt,
    author = {Ares, F{\"e}anor Reuben and Hindmarsh, Mark and Hoyos, Carlos and Jokela, Niko},
    title = "{Gravitational waves from a holographic phase transition}",
    eprint = "2011.12878",
    archivePrefix = "arXiv",
    primaryClass = "hep-th",
    reportNumber = "HIP-2020-31/TH, Sussex-94886",
    doi = "10.1007/JHEP04(2021)100",
    journal = "JHEP",
    volume = "21",
    pages = "100",
    year = "2020"
}

@article{Ares:2021ntv,
    author = {Ares, F{\"e}anor Reuben and Henriksson, Oscar and Hindmarsh, Mark and Hoyos, Carlos and Jokela, Niko},
    title = "{Effective actions and bubble nucleation from holography}",
    eprint = "2109.13784",
    archivePrefix = "arXiv",
    primaryClass = "hep-th",
    reportNumber = "HIP-2021-30/TH",
    doi = "10.1103/PhysRevD.105.066020",
    journal = "Phys. Rev. D",
    volume = "105",
    number = "6",
    pages = "066020",
    year = "2022"
}

@article{Ares:2021nap,
    author = {Ares, F{\"e}anor Reuben and Henriksson, Oscar and Hindmarsh, Mark and Hoyos, Carlos and Jokela, Niko},
    title = "{Gravitational Waves at Strong Coupling from an Effective Action}",
    eprint = "2110.14442",
    archivePrefix = "arXiv",
    primaryClass = "hep-th",
    reportNumber = "HIP-2021-24/TH",
    doi = "10.1103/PhysRevLett.128.131101",
    journal = "Phys. Rev. Lett.",
    volume = "128",
    number = "13",
    pages = "131101",
    year = "2022"
}

@article{Bea:2021zol,
    author = "Bea, Yago and Casalderrey-Solana, Jorge and Giannakopoulos, Thanasis and Jansen, Aron and Krippendorf, Sven and Mateos, David and Sanchez-Garitaonandia, Mikel and Zilh{\~a}o, Miguel",
    title = "{Spinodal Gravitational Waves}",
    eprint = "2112.15478",
    archivePrefix = "arXiv",
    primaryClass = "hep-th",
    reportNumber = "LMU-ASC 59/21",
    doi = "10.1007/JHEP11(2025)093",
    journal = "JHEP",
    volume = "11",
    pages = "093",
    year = "2025"
}

@article{Buchel:2021yay,
    author = "Buchel, Alex",
    title = "{A bestiary of black holes on the conifold with fluxes}",
    eprint = "2103.15188",
    archivePrefix = "arXiv",
    primaryClass = "hep-th",
    doi = "10.1007/JHEP06(2021)102",
    journal = "JHEP",
    volume = "06",
    pages = "102",
    year = "2021"
}

@article{Buchel:2018bzp,
    author = "Buchel, Alex",
    title = "{Klebanov-Strassler black hole}",
    eprint = "1809.08484",
    archivePrefix = "arXiv",
    primaryClass = "hep-th",
    doi = "10.1007/JHEP01(2019)207",
    journal = "JHEP",
    volume = "01",
    pages = "207",
    year = "2019"
}

@article{Buchel:2024phy,
    author = "Buchel, Alex",
    title = "{Fluxification and scalarization of the conifold black holes}",
    eprint = "2411.15950",
    archivePrefix = "arXiv",
    primaryClass = "hep-th",
    doi = "10.1103/PhysRevD.111.046010",
    journal = "Phys. Rev. D",
    volume = "111",
    number = "4",
    pages = "046010",
    year = "2025"
}

@article{LISA:2017pwj,
    author = "Amaro-Seoane, Pau and others",
    collaboration = "LISA",
    title = "{Laser Interferometer Space Antenna}",
    eprint = "1702.00786",
    archivePrefix = "arXiv",
    primaryClass = "astro-ph.IM",
    month = "2",
    year = "2017"
}

@article{Baker:2019nia,
    author = "Baker, John and others",
    title = "{The Laser Interferometer Space Antenna: Unveiling the Millihertz Gravitational Wave Sky}",
    eprint = "1907.06482",
    archivePrefix = "arXiv",
    primaryClass = "astro-ph.IM",
    reportNumber = "FERMILAB-PUB-19-436-A",
    month = "7",
    year = "2019"
}

@article{Seto:2001qf,
    author = "Seto, Naoki and Kawamura, Seiji and Nakamura, Takashi",
    title = "{Possibility of direct measurement of the acceleration of the universe using 0.1-Hz band laser interferometer gravitational wave antenna in space}",
    eprint = "astro-ph/0108011",
    archivePrefix = "arXiv",
    doi = "10.1103/PhysRevLett.87.221103",
    journal = "Phys. Rev. Lett.",
    volume = "87",
    pages = "221103",
    year = "2001"
}

@article{Kawamura:2011zz,
    author = "Kawamura, Seiji and others",
    editor = "Buchman, Sasha and Sun, Ke-Xun",
    title = "{The Japanese space gravitational wave antenna: DECIGO}",
    doi = "10.1088/0264-9381/28/9/094011",
    journal = "Class. Quant. Grav.",
    volume = "28",
    pages = "094011",
    year = "2011"
}

@article{Yagi:2011wg,
    author = "Yagi, Kent and Seto, Naoki",
    title = "{Detector configuration of DECIGO/BBO and identification of cosmological neutron-star binaries}",
    eprint = "1101.3940",
    archivePrefix = "arXiv",
    primaryClass = "astro-ph.CO",
    doi = "10.1103/PhysRevD.83.044011",
    journal = "Phys. Rev. D",
    volume = "83",
    pages = "044011",
    year = "2011",
    note = "[Erratum: Phys.Rev.D 95, 109901 (2017)]"
}

@article{Isoyama:2018rjb,
    author = "Isoyama, Soichiro and Nakano, Hiroyuki and Nakamura, Takashi",
    title = "{Multiband Gravitational-Wave Astronomy: Observing binary inspirals with a decihertz detector, B-DECIGO}",
    eprint = "1802.06977",
    archivePrefix = "arXiv",
    primaryClass = "gr-qc",
    doi = "10.1093/ptep/pty078",
    journal = "PTEP",
    volume = "2018",
    number = "7",
    pages = "073E01",
    year = "2018"
}

@article{Crowder:2005nr,
    author = "Crowder, Jeff and Cornish, Neil J.",
    title = "{Beyond LISA: Exploring future gravitational wave missions}",
    eprint = "gr-qc/0506015",
    archivePrefix = "arXiv",
    doi = "10.1103/PhysRevD.72.083005",
    journal = "Phys. Rev. D",
    volume = "72",
    pages = "083005",
    year = "2005"
}

@article{Corbin:2005ny,
    author = "Corbin, Vincent and Cornish, Neil J.",
    title = "{Detecting the cosmic gravitational wave background with the big bang observer}",
    eprint = "gr-qc/0512039",
    archivePrefix = "arXiv",
    doi = "10.1088/0264-9381/23/7/014",
    journal = "Class. Quant. Grav.",
    volume = "23",
    pages = "2435--2446",
    year = "2006"
}

@article{Harry:2006fi,
    author = "Harry, G. M. and Fritschel, P. and Shaddock, D. A. and Folkner, W. and Phinney, E. S.",
    title = "{Laser interferometry for the big bang observer}",
    doi = "10.1088/0264-9381/23/15/008",
    journal = "Class. Quant. Grav.",
    volume = "23",
    pages = "4887--4894",
    year = "2006",
    note = "[Erratum: Class.Quant.Grav. 23, 7361 (2006)]"
}

@article{LIGOScientific:2016wof,
    author = "Abbott, Benjamin P and others",
    collaboration = "LIGO Scientific",
    title = "{Exploring the Sensitivity of Next Generation Gravitational Wave Detectors}",
    eprint = "1607.08697",
    archivePrefix = "arXiv",
    primaryClass = "astro-ph.IM",
    reportNumber = "LIGO-P1600143",
    doi = "10.1088/1361-6382/aa51f4",
    journal = "Class. Quant. Grav.",
    volume = "34",
    number = "4",
    pages = "044001",
    year = "2017"
}

@article{Reitze:2019iox,
    author = "Reitze, David and others",
    title = "{Cosmic Explorer: The U.S. Contribution to Gravitational-Wave Astronomy beyond LIGO}",
    eprint = "1907.04833",
    archivePrefix = "arXiv",
    primaryClass = "astro-ph.IM",
    reportNumber = "LIGO-P1900316",
    journal = "Bull. Am. Astron. Soc.",
    volume = "51",
    number = "7",
    pages = "035",
    year = "2019"
}

@article{Punturo:2010zz,
    author = "Punturo, M. and others",
    editor = "Ricci, Fulvio",
    title = "{The Einstein Telescope: A third-generation gravitational wave observatory}",
    doi = "10.1088/0264-9381/27/19/194002",
    journal = "Class. Quant. Grav.",
    volume = "27",
    pages = "194002",
    year = "2010"
}

@article{Hild:2010id,
    author = "Hild, S. and others",
    title = "{Sensitivity Studies for Third-Generation Gravitational Wave Observatories}",
    eprint = "1012.0908",
    archivePrefix = "arXiv",
    primaryClass = "gr-qc",
    doi = "10.1088/0264-9381/28/9/094013",
    journal = "Class. Quant. Grav.",
    volume = "28",
    pages = "094013",
    year = "2011"
}

@article{Sathyaprakash:2012jk,
    author = "Sathyaprakash, B. and others",
    editor = "Hannam, Mark and Sutton, Patrick and Hild, Stefan and van den Broeck, Chris",
    title = "{Scientific Objectives of Einstein Telescope}",
    eprint = "1206.0331",
    archivePrefix = "arXiv",
    primaryClass = "gr-qc",
    doi = "10.1088/0264-9381/29/12/124013",
    journal = "Class. Quant. Grav.",
    volume = "29",
    pages = "124013",
    year = "2012",
    note = "[Erratum: Class.Quant.Grav. 30, 079501 (2013)]"
}

@article{ET:2019dnz,
    author = "Maggiore, Michele and others",
    collaboration = "ET",
    title = "{Science Case for the Einstein Telescope}",
    eprint = "1912.02622",
    archivePrefix = "arXiv",
    primaryClass = "astro-ph.CO",
    doi = "10.1088/1475-7516/2020/03/050",
    journal = "JCAP",
    volume = "03",
    pages = "050",
    year = "2020"
}

@article{AEDGE:2019nxb,
    author = "El-Neaj, Yousef Abou and others",
    collaboration = "AEDGE",
    title = "{AEDGE: Atomic Experiment for Dark Matter and Gravity Exploration in Space}",
    eprint = "1908.00802",
    archivePrefix = "arXiv",
    primaryClass = "gr-qc",
    reportNumber = "KCL-PH-TH/2019-65, CERN-TH-2019-126",
    doi = "10.1140/epjqt/s40507-020-0080-0",
    journal = "EPJ Quant. Technol.",
    volume = "7",
    pages = "6",
    year = "2020"
}

@article{Badurina:2019hst,
    author = "Badurina, L. and others",
    title = "{AION: An Atom Interferometer Observatory and Network}",
    eprint = "1911.11755",
    archivePrefix = "arXiv",
    primaryClass = "astro-ph.CO",
    reportNumber = "AION-2019-001, CERN-TH-2019-199",
    doi = "10.1088/1475-7516/2020/05/011",
    journal = "JCAP",
    volume = "05",
    pages = "011",
    year = "2020"
}

@article{Caprini:2024hue,
    author = "Caprini, Chiara and Jinno, Ryusuke and Lewicki, Marek and Madge, Eric and Merchand, Marco and Nardini, Germano and Pieroni, Mauro and Roper Pol, Alberto and Vaskonen, Ville",
    collaboration = "LISA Cosmology Working Group",
    title = "{Gravitational waves from first-order phase transitions in LISA: reconstruction pipeline and physics interpretation}",
    eprint = "2403.03723",
    archivePrefix = "arXiv",
    primaryClass = "astro-ph.CO",
    reportNumber = "LISA-COSWG-24-01, CERN-TH-2024-029",
    doi = "10.1088/1475-7516/2024/10/020",
    journal = "JCAP",
    volume = "10",
    pages = "020",
    year = "2024"
}

@article{Schmitz:2020syl,
    author = "Schmitz, Kai",
    title = "{New Sensitivity Curves for Gravitational-Wave Signals from Cosmological Phase Transitions}",
    eprint = "2002.04615",
    archivePrefix = "arXiv",
    primaryClass = "hep-ph",
    reportNumber = "CERN-TH-2020-018",
    doi = "10.1007/JHEP01(2021)097",
    journal = "JHEP",
    volume = "01",
    pages = "097",
    year = "2021"
}

@article{Laurent:2022jrs,
    author = "Laurent, Benoit and Cline, James M.",
    title = "{First principles determination of bubble wall velocity}",
    eprint = "2204.13120",
    archivePrefix = "arXiv",
    primaryClass = "hep-ph",
    doi = "10.1103/PhysRevD.106.023501",
    journal = "Phys. Rev. D",
    volume = "106",
    number = "2",
    pages = "023501",
    year = "2022"
}

@article{Bigazzi:2021ucw,
    author = "Bigazzi, Francesco and Caddeo, Alessio and Canneti, Tommaso and Cotrone, Aldo L.",
    title = "{Bubble wall velocity at strong coupling}",
    eprint = "2104.12817",
    archivePrefix = "arXiv",
    primaryClass = "hep-ph",
    doi = "10.1007/JHEP08(2021)090",
    journal = "JHEP",
    volume = "08",
    pages = "090",
    year = "2021"
}

@article{Bea:2021zsu,
    author = "Bea, Yago and Casalderrey-Solana, Jorge and Giannakopoulos, Thanasis and Mateos, David and Sanchez-Garitaonandia, Mikel and Zilh{\~a}o, Miguel",
    title = "{Bubble wall velocity from holography}",
    eprint = "2104.05708",
    archivePrefix = "arXiv",
    primaryClass = "hep-th",
    doi = "10.1103/PhysRevD.104.L121903",
    journal = "Phys. Rev. D",
    volume = "104",
    number = "12",
    pages = "L121903",
    year = "2021"
}

@article{Falkowski:2008fz,
    author = "Falkowski, Adam and Perez-Victoria, Manuel",
    title = "{Electroweak Breaking on a Soft Wall}",
    eprint = "0806.1737",
    archivePrefix = "arXiv",
    primaryClass = "hep-ph",
    doi = "10.1088/1126-6708/2008/12/107",
    journal = "JHEP",
    volume = "12",
    pages = "107",
    year = "2008"
}

@article{Falkowski:2008yr,
    author = "Falkowski, Adam and Perez-Victoria, Manuel",
    title = "{Holographic Unhiggs}",
    eprint = "0810.4940",
    archivePrefix = "arXiv",
    primaryClass = "hep-ph",
    doi = "10.1103/PhysRevD.79.035005",
    journal = "Phys. Rev. D",
    volume = "79",
    pages = "035005",
    year = "2009"
}

@article{Cabrer:2009we,
    author = "Cabrer, Joan A. and von Gersdorff, Gero and Quiros, Mariano",
    title = "{Soft-Wall Stabilization}",
    eprint = "0907.5361",
    archivePrefix = "arXiv",
    primaryClass = "hep-ph",
    reportNumber = "CERN-PH-TH-2009-141",
    doi = "10.1088/1367-2630/12/7/075012",
    journal = "New J. Phys.",
    volume = "12",
    pages = "075012",
    year = "2010"
}

@article{Bellazzini:2015cgj,
    author = "Bellazzini, Brando and Cs{\'a}ki, Csaba and Hubisz, Jay and Lee, Seung J. and Serra, Javi and Terning, John",
    title = "{Quantum Critical Higgs}",
    eprint = "1511.08218",
    archivePrefix = "arXiv",
    primaryClass = "hep-ph",
    reportNumber = "CERN-PH-TH-2015-275, Saclay-t15-206",
    doi = "10.1103/PhysRevX.6.041050",
    journal = "Phys. Rev. X",
    volume = "6",
    number = "4",
    pages = "041050",
    year = "2016"
}

@article{Csaki:2018kxb,
    author = "Cs{\'a}ki, Csaba and Lee, Gabriel and Lee, Seung J. and Lombardo, Salvator and Telem, Ofri",
    title = "{Continuum Naturalness}",
    eprint = "1811.06019",
    archivePrefix = "arXiv",
    primaryClass = "hep-ph",
    doi = "10.1007/JHEP03(2019)142",
    journal = "JHEP",
    volume = "03",
    pages = "142",
    year = "2019"
}

@article{Megias:2019vdb,
    author = "Meg{\'\i}as, Eugenio and Quir{\'o}s, Mariano",
    title = "{Gapped Continuum Kaluza-Klein spectrum}",
    eprint = "1905.07364",
    archivePrefix = "arXiv",
    primaryClass = "hep-ph",
    reportNumber = "UAB-FT-779",
    doi = "10.1007/JHEP08(2019)166",
    journal = "JHEP",
    volume = "08",
    pages = "166",
    year = "2019"
}

@article{Megias:2021mgj,
    author = "Megias, Eugenio and Quiros, Mariano",
    title = "{The Continuum Linear Dilaton}",
    eprint = "2104.10260",
    archivePrefix = "arXiv",
    primaryClass = "hep-ph",
    doi = "10.5506/APhysPolB.52.711",
    journal = "Acta Phys. Polon. B",
    volume = "52",
    number = "6-7",
    pages = "711",
    year = "2021"
}

@article{Csaki:2021gfm,
    author = "Cs{\'a}ki, Csaba and Hong, Sungwoo and Kurup, Gowri and Lee, Seung J. and Perelstein, Maxim and Xue, Wei",
    title = "{Continuum dark matter}",
    eprint = "2105.07035",
    archivePrefix = "arXiv",
    primaryClass = "hep-ph",
    doi = "10.1103/PhysRevD.105.035025",
    journal = "Phys. Rev. D",
    volume = "105",
    number = "3",
    pages = "035025",
    year = "2022"
}

@article{Csaki:2021xpy,
    author = "Cs{\'a}ki, Csaba and Hong, Sungwoo and Kurup, Gowri and Lee, Seung J. and Perelstein, Maxim and Xue, Wei",
    title = "{Z-Portal Continuum Dark Matter}",
    eprint = "2105.14023",
    archivePrefix = "arXiv",
    primaryClass = "hep-ph",
    doi = "10.1103/PhysRevLett.128.081807",
    journal = "Phys. Rev. Lett.",
    volume = "128",
    number = "8",
    pages = "081807",
    year = "2022"
}

@article{Csaki:2022lnq,
    author = "Csaki, Csaba and Ismail, Ameen and Lee, Seung J.",
    title = "{The continuum dark matter zoo}",
    eprint = "2210.16326",
    archivePrefix = "arXiv",
    primaryClass = "hep-ph",
    doi = "10.1007/JHEP02(2023)053",
    journal = "JHEP",
    volume = "02",
    pages = "053",
    year = "2023"
}

@article{Fichet:2022xol,
    author = "Fichet, Sylvain and Megias, Eugenio and Quiros, Mariano",
    title = "{Cosmological dark matter from a bulk black hole}",
    eprint = "2212.13268",
    archivePrefix = "arXiv",
    primaryClass = "hep-ph",
    doi = "10.1103/PhysRevD.107.115014",
    journal = "Phys. Rev. D",
    volume = "107",
    number = "11",
    pages = "115014",
    year = "2023"
}

@article{Fichet:2023xbu,
    author = "Fichet, Sylvain and Megias, Eugenio and Quiros, Mariano",
    title = "{Holography of linear dilaton spacetimes from the bottom up}",
    eprint = "2309.02489",
    archivePrefix = "arXiv",
    primaryClass = "hep-th",
    doi = "10.1103/PhysRevD.109.106011",
    journal = "Phys. Rev. D",
    volume = "109",
    number = "10",
    pages = "106011",
    year = "2024"
}

@article{Fichet:2023dju,
    author = "Fichet, Sylvain and Megias, Eugenio and Quiros, Mariano",
    title = "{Holographic fluids from 5D dilaton gravity}",
    eprint = "2311.14233",
    archivePrefix = "arXiv",
    primaryClass = "hep-th",
    doi = "10.1007/JHEP08(2024)077",
    journal = "JHEP",
    volume = "08",
    pages = "077",
    year = "2024"
}

@article{Ferrante:2023fpx,
    author = "Ferrante, Steven and Lee, Seung J. and Perelstein, Maxim",
    title = "{Collider signatures of near-continuum dark matter}",
    eprint = "2306.13009",
    archivePrefix = "arXiv",
    primaryClass = "hep-ph",
    doi = "10.1007/JHEP05(2024)215",
    journal = "JHEP",
    volume = "05",
    pages = "215",
    year = "2024"
}

@article{Ferrante:2025ofe,
    author = "Ferrante, Steven and Luo, Lillian and Perelstein, Maxim and Youn, Taewook",
    title = "{Collider Searches for Near-Continuum Dark Matter}",
    eprint = "2510.17989",
    archivePrefix = "arXiv",
    primaryClass = "hep-ph",
    month = "10",
    year = "2025"
}

@article{Gursoy:2007cb,
    author = "Gursoy, U. and Kiritsis, E.",
    title = "{Exploring improved holographic theories for QCD: Part I}",
    eprint = "0707.1324",
    archivePrefix = "arXiv",
    primaryClass = "hep-th",
    reportNumber = "CPHT-RR027-0507",
    doi = "10.1088/1126-6708/2008/02/032",
    journal = "JHEP",
    volume = "02",
    pages = "032",
    year = "2008"
}

@article{Gursoy:2007er,
    author = "Gursoy, U. and Kiritsis, E. and Nitti, F.",
    title = "{Exploring improved holographic theories for QCD: Part II}",
    eprint = "0707.1349",
    archivePrefix = "arXiv",
    primaryClass = "hep-th",
    reportNumber = "CPHT-RR028-0507",
    doi = "10.1088/1126-6708/2008/02/019",
    journal = "JHEP",
    volume = "02",
    pages = "019",
    year = "2008"
}

@article{DeWolfe:1999cp,
    author = "DeWolfe, O. and Freedman, D. Z. and Gubser, S. S. and Karch, A.",
    title = "{Modeling the fifth-dimension with scalars and gravity}",
    eprint = "hep-th/9909134",
    archivePrefix = "arXiv",
    reportNumber = "HUTP-99-A048, MIT-CTP-2903",
    doi = "10.1103/PhysRevD.62.046008",
    journal = "Phys. Rev. D",
    volume = "62",
    pages = "046008",
    year = "2000"
}

@article{Csaki:2000zn,
    author = "Csaki, Csaba and Graesser, Michael L. and Kribs, Graham D.",
    title = "{Radion dynamics and electroweak physics}",
    eprint = "hep-th/0008151",
    archivePrefix = "arXiv",
    reportNumber = "SCIPP-00-27",
    doi = "10.1103/PhysRevD.63.065002",
    journal = "Phys. Rev. D",
    volume = "63",
    pages = "065002",
    year = "2001"
}

@article{Gubser:2000nd,
    author = "Gubser, Steven S.",
    title = "{Curvature singularities: The Good, the bad, and the naked}",
    eprint = "hep-th/0002160",
    archivePrefix = "arXiv",
    reportNumber = "PUPT-1916",
    doi = "10.4310/ATMP.2000.v4.n3.a6",
    journal = "Adv. Theor. Math. Phys.",
    volume = "4",
    pages = "679--745",
    year = "2000"
}

@article{Guada:2020xnz,
    author = "Guada, Victor and Nemev{\v{s}}ek, Miha and Pintar, Matev{\v{z}}",
    title = "{FindBounce: Package for multi-field bounce actions}",
    eprint = "2002.00881",
    archivePrefix = "arXiv",
    primaryClass = "hep-ph",
    doi = "10.1016/j.cpc.2020.107480",
    journal = "Comput. Phys. Commun.",
    volume = "256",
    pages = "107480",
    year = "2020"
}

@article{Coleman:1977py,
    author = "Coleman, Sidney R.",
    title = "{The Fate of the False Vacuum. 1. Semiclassical Theory}",
    reportNumber = "HUTP-77-A004",
    doi = "10.1103/PhysRevD.16.1248",
    journal = "Phys. Rev. D",
    volume = "15",
    pages = "2929--2936",
    year = "1977",
    note = "[Erratum: Phys.Rev.D 16, 1248 (1977)]"
}

@article{Linde:1981zj,
    author = "Linde, Andrei D.",
    title = "{Decay of the False Vacuum at Finite Temperature}",
    reportNumber = "LEBEDEV-81-265",
    doi = "10.1016/0550-3213(83)90072-X",
    journal = "Nucl. Phys. B",
    volume = "216",
    pages = "421",
    year = "1983",
    note = "[Erratum: Nucl.Phys.B 223, 544 (1983)]"
}

@article{Aharony:2005bm,
    author = "Aharony, Ofer and Minwalla, Shiraz and Wiseman, Toby",
    title = "{Plasma-balls in large N gauge theories and localized black holes}",
    eprint = "hep-th/0507219",
    archivePrefix = "arXiv",
    reportNumber = "WIS-18-05-JUL-DPP, HUTP-05-A0035",
    doi = "10.1088/0264-9381/23/7/001",
    journal = "Class. Quant. Grav.",
    volume = "23",
    pages = "2171--2210",
    year = "2006"
}

@article{Bea:2022mfb,
    author = "Bea, Yago and Casalderrey-Solana, Jorge and Giannakopoulos, Thanasis and Jansen, Aron and Mateos, David and Sanchez-Garitaonandia, Mikel and Zilh{\~a}o, Miguel",
    title = "{Holographic bubbles with Jecco: expanding, collapsing and critical}",
    eprint = "2202.10503",
    archivePrefix = "arXiv",
    primaryClass = "hep-th",
    doi = "10.1007/JHEP09(2022)008",
    journal = "JHEP",
    volume = "09",
    pages = "008",
    year = "2022",
    note = "[Erratum: JHEP 03, 225 (2023)]"
}

@article{Bigazzi:2020phm,
    author = "Bigazzi, Francesco and Caddeo, Alessio and Cotrone, Aldo L. and Paredes, Angel",
    title = "{Fate of false vacua in holographic first-order phase transitions}",
    eprint = "2008.02579",
    archivePrefix = "arXiv",
    primaryClass = "hep-th",
    doi = "10.1007/JHEP12(2020)200",
    journal = "JHEP",
    volume = "12",
    pages = "200",
    year = "2020"
}

@article{Csaki:1999mp,
    author = "Csaki, Csaba and Graesser, Michael and Randall, Lisa and Terning, John",
    title = "{Cosmology of brane models with radion stabilization}",
    eprint = "hep-ph/9911406",
    archivePrefix = "arXiv",
    reportNumber = "SCIPP-99-49, HUTP-A061, NSF-ITP-99-130",
    doi = "10.1103/PhysRevD.62.045015",
    journal = "Phys. Rev. D",
    volume = "62",
    pages = "045015",
    year = "2000"
}

@article{Goldberger:1999un,
    author = "Goldberger, Walter D. and Wise, Mark B.",
    title = "{Phenomenology of a stabilized modulus}",
    eprint = "hep-ph/9911457",
    archivePrefix = "arXiv",
    reportNumber = "CALT-68-2250",
    doi = "10.1016/S0370-2693(00)00099-X",
    journal = "Phys. Lett. B",
    volume = "475",
    pages = "275--279",
    year = "2000"
}

@article{Skenderis:2002wp,
    author = "Skenderis, Kostas",
    editor = "de Wit, B. and Vandoren, S.",
    title = "{Lecture notes on holographic renormalization}",
    eprint = "hep-th/0209067",
    archivePrefix = "arXiv",
    reportNumber = "PUTP-2047",
    doi = "10.1088/0264-9381/19/22/306",
    journal = "Class. Quant. Grav.",
    volume = "19",
    pages = "5849--5876",
    year = "2002"
}

@article{Karch:2006pv,
    author = "Karch, Andreas and Katz, Emanuel and Son, Dam T. and Stephanov, Mikhail A.",
    title = "{Linear confinement and AdS/QCD}",
    eprint = "hep-ph/0602229",
    archivePrefix = "arXiv",
    reportNumber = "BUHEP-06-02, INT-PUB-06-04",
    doi = "10.1103/PhysRevD.74.015005",
    journal = "Phys. Rev. D",
    volume = "74",
    pages = "015005",
    year = "2006"
}

\end{document}